\documentclass[sigconf]{acmart}
\usepackage{comment}
\usepackage{xspace}
\usepackage{enumitem}
\usepackage[ruled,linesnumbered,noend]{algorithm2e}
\usepackage{subcaption}
\usepackage{xcolor}
\usepackage{multirow}
\usepackage{booktabs}
\usepackage{url}
\usepackage{adjustbox}
\usepackage{dsfont}

\newcommand{\FreSCo}{\textsc{FreSCo}\xspace}
\newcommand{\Benson}{\textsc{B-Exact}\xspace}
\newcommand{\coauthDBLP}{\textt{cD}\xspace}
\newcommand{\coauthMAGGeology}{\textt{cMG}\xspace}
\newcommand{\coauthMAGHistory}{\textt{cMH}\xspace}

\newcommand{\contacthighschool}{\textt{chs}\xspace}
\newcommand{\contactprimaryschool}{\textt{cps}\xspace}

\newcommand{\emailEu}{\textt{eEu}\xspace}
\newcommand{\emailEnron}{\textt{eEn}\xspace}

\newcommand{\tagsaskubuntu}{\textt{taau}\xspace}
\newcommand{\tagsstackoverflow}{\textt{taso}\xspace}
\newcommand{\threadsaskubuntu}{\textt{thau}\xspace}
\newcommand{\threadsmathsx}{\textt{thms}\xspace}
\newcommand{\threadsstackoverflow}{\textt{thso}\xspace}
\newcommand{\kmean}{k-means++\xspace}

\newcommand{\rome}[1]{\uppercase\expandafter{\romannumeral #1\relax}}

\newcommand{\smallsection}[1]{\noindent\underline{\smash{\textbf{#1:}}}}

\newcommand{\textt}[1]{\scalebox{1.0}{\texttt{#1}}}

\newtheorem{theorem}{Theorem}
\newtheorem{lemma}{Lemma}
\newtheorem{definition}{Definition}
\newtheorem{problem}{Problem}

\SetKwComment{Comment}{$\triangleright$\ }{}

\SetCommentSty{mycommfont}
\SetAlFnt{\small}

\newcommand{\bus}[1]{\textbf{\underline{\smash{#1}}}}
\newcommand{\ours}{SC3\xspace}
\newcommand{\OURS}{\bus{S}implet \bus{C}ounting using \bus{C}olor \bus{C}oding\xspace}

\AtBeginDocument{%
  }

\copyrightyear{2023}
\acmYear{2023}
\setcopyright{acmlicensed}\acmConference[WWW '23]{Proceedings of the ACM Web Conference 2023}{May 1--5, 2023}{Austin, TX, USA}
\acmBooktitle{Proceedings of the ACM Web Conference 2023 (WWW '23), May 1--5, 2023, Austin, TX, USA}
\acmPrice{15.00}
\acmDOI{10.1145/3543507.3583332}
\acmISBN{978-1-4503-9416-1/23/04}

\begin{document}

\title{Characterization of Simplicial Complexes by Counting Simplets Beyond Four Nodes}

\settopmatter{authorsperrow=4}
\author{Hyunju Kim}
\affiliation{%
  \institution{Kim Jaechul Graduate School of AI, KAIST}
  \city{Seoul}
  \country{South Korea}
}
\email{hyunju.kim@kaist.ac.kr}

\author{Jihoon Ko}
\affiliation{%
  \institution{Kim Jaechul Graduate School of AI, KAIST}
  \city{Seoul}
  \country{South Korea}
}
\email{jihoonko@kaist.ac.kr}

\author{Fanchen Bu}
\affiliation{%
  \institution{School of Electrical Engineering, KAIST}
  \city{Daejeon}
  \country{South Korea}
}
\email{boqvezen97@kaist.ac.kr}

\author{Kijung Shin}
\affiliation{%
  \institution{Kim Jaechul Graduate School of AI, KAIST}
  \city{Seoul}
  \country{South Korea}
}
\email{kijungs@kaist.ac.kr}

\begin{abstract}
Simplicial complexes are higher-order combinatorial structures which have been used to represent real-world complex systems.
In this paper, we concentrate on the local patterns in simplicial complexes called \textit{simplets}, a generalization of graphlets.
We formulate the problem of counting simplets of a given size in a given simplicial complex.
For this problem, we extend a sampling algorithm based on color coding from graphs to simplicial complexes, with essential technical novelty.
We theoretically analyze our proposed algorithm named \ours, showing its correctness, unbiasedness, convergence, and time/space complexity. %
Through the extensive experiments on sixteen real-world datasets, we show the superiority of \ours in terms of  accuracy, speed, and scalability, compared to the baseline methods.
Finally, we use the counts given by \ours for simplicial complex analysis, especially for characterization, which is further used for simplicial complex clustering, where \ours shows a strong ability of characterization with domain-based similarity.
    
\end{abstract}

\ccsdesc[500]{Information systems~Data mining}
\ccsdesc[500]{Information systems~Social networks}
\keywords{Simplicial Complex, Simplet, Counting, 
Characterization}

\maketitle

\section{Introduction}
\label{sec:intro}
In many real-world systems, group relations involving more than two entities exist, which cannot be fully represented by pairwise graphs.
For example, for co-authorship relations~\cite{sinha2015overview}, a single publication is possibly done by more than two authors;
for email systems~\cite{leskovec2007graph}, the recipients of an email can be more than two.
Therefore, hypergraphs, where an edge may contain more than two nodes,
are able to naturally represent group relations involving more than two entities, and thus used to model such systems.

\begin{figure}
    \centering
    \includegraphics[width=0.968\linewidth]{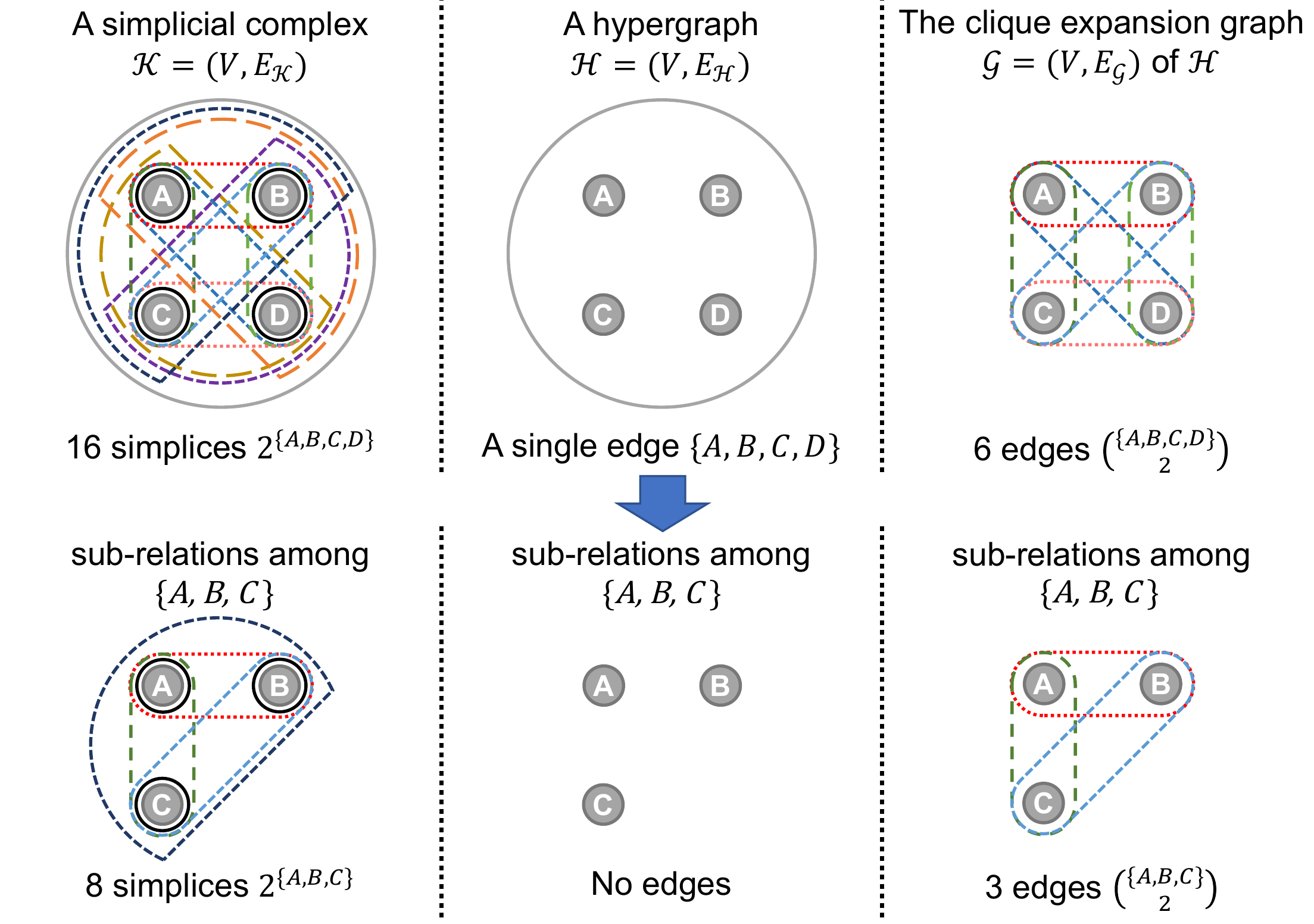}
    \caption{A simplicial complex (left), a hypergraph (middle), and a clique expansion graph (right) representing the same group interaction among four entities. The hypergraph fails to capture interaction involving only a subset of the entities while the clique expansion graph fails to represent interactions among groups of size more than two.}
    \label{fig:intro}
\end{figure}

In spite of the representative power of hypergraphs,
a potential problem of modeling real-world systems as hypergraphs is that, for a group relation, although all of its subset relations naturally exist, usually only the largest group relation is represented in the hypergraph as an edge.
For example, when we have a group relation involving four entities $a, b, c, d$, usually only a single edge $\{a, b, c, d\}$ exists in the corresponding hypergraph, {as shown in Figure~\ref{fig:intro}}, which overlooks, e.g., the relations $\{a, b, c\}$,  $\{a, b, d\}$, $\{a, b\}$ and so on.,
making it hard to capture local patterns (cf. the clique expansion graph where group interactions of size more than two are ignored).

In order to address this problem,
we may use simplicial complexes~\cite{milnor1957geometric,street1987algebra,jonsson2007simplicial} (SCs).
With the \textit{downward closure} property, for each edge in an SC, all of its subsets are also included in the SC, as shown in Figure~\ref{fig:intro}.
Compared to hypergraph-based modeling, SC-based modeling has advantages in applications and connections to geometry and algebra~\cite{schaub2021signal},
and in practice, we do not need additional space for the SC-based modeling since we can store the same set of edges as in the hypergraph while all the subset edges are implicitly included.
Due to these merits, SCs have been used to model real-world complex systems~\cite{salnikov2018simplicial, benson2018simplicial} of communication~\cite{wang2020social}, epidemic spreading~\cite{li2021contagion}, or social contagion~\cite{iacopini2019simplicial}.
See also~\cite{torres2021and} for comprehensive comparisons between SCs and (hyper)graphs.

One of the benefits of modeling real-world systems as abstract structures mentioned above is that it makes studying the patterns~\cite{cheng2014mining} within the systems easier, especially the local patterns~\cite{knobbe2008local}.
An important and widely-used example on pairwise graphs is graphlets~\cite{milo2002network,prvzulj2004modeling} which 
describe the pattern of the interactions among several nodes.
In typical usage of graphlets, the occurrences of each graphlet are counted in the input graph~\cite{prvzulj2007biological,milenkovic2010optimal}, and the counts are used to measure the similarity between graphs~\cite{shervashidze2009efficient}, detect anomalies~\cite{harshaw2016graphprints}, or detect communities~\cite{zhu2021community}.

In this paper, we consider the problem of counting \textit{simplets}, the counterpart concept of graphlets on SCs.
Similar to graphlets, simplets describe the patterns of the simplices formed by each group of nodes.
The concept of simplets and the problem of counting simplets are mentioned for the first time in~\cite{preti2022fresco}.
However, in~\cite{preti2022fresco}, instead of directly counting the occurrences of simplets, a surrogate measure is proposed and used to indirectly estimate the occurrences, and actually only the problem of computing the surrogate measure is studied.
The authors do so mainly due to the theoretical hardness of directly counting the occurrences of simplets.
In this work, we study the problem of directly counting the simplest. %

Many techniques have been proposed for counting graphlets~\cite{hovcevar2014combinatorial, ahmed2015efficient, pinar2017escape, wang2017moss}.
Recently, graphlet-counting methods~\cite{bressan2017counting, bressan2018motif, bressan2019motivo} based on color coding (CC)~\cite{alon1995color} are proposed, especially for the graphlets of sizes more than five. 
For simplet counting, we propose \textbf{\ours} (\OURS),
an algorithm using CC-based sampling,
where we adapt the algorithms for graphlet counting in~\cite{bressan2017counting, bressan2018motif, bressan2019motivo} with technical improvement, to deal with the intrinsic hardness of simplet counting on set enumeration, set isomorphism check, etc.
Specifically, given an SC and a specific size of simplets, we first use a standard CC process consisting of two steps (building and sampling) to sample a candidate set of node sets.
After that, we obtain the maximal simplices in the subcomplex induced on each sampled set via the scanning step,
and finally match each group of maximal simplices with a simplet by considering all the permutations of each simplet.
We also theoretically prove the correctness, unbiasedness, convergence, and time/space complexity of \ours. 

Through extensive experiments on sixteen real-world datasets, we show the empirical correctness, convergence, and high speed of \ours.
Specifically, for size-4 simplets, with $100,000$ samples, the counts given by \ours have a normalized error (normalized by the ground truth total number of simplets) lower than $5\%$ on all the datasets.
Regarding the speed, \ours is up to $10,000$ times faster than an exact method.
\ours also shows excellent performance on counting size-$5$ simplest. %

We also use the counts obtained by \ours for characterizing SCs.
Specifically, we normalize the count of each simplet by comparing it with the count in a null model and form a characteristic vector by combining the counts of all simplets.
We show that \ours has the highest characterization power compared to the baseline methods.
Furthermore, we apply \kmean~\cite{arthur2006k} to the characteristic vectors for clustering SCs, where perfect clustering results are shown when we use \ours on size-5 simplets.

In short, our contributions are four-fold:
\begin{itemize}[leftmargin=*]
    \item \textbf{New problem.} To the best of our knowledge, we formulate and study the problem of directly counting simplets in a given SC for the first time, especially for the simplets of sizes more than four.
    \item \textbf{Algorithm.} We propose \ours for the simplet counting problem using color-coding-based sampling, and prove its correctness, unbiasedness, convergence, and time/space complexity.
    \item \textbf{Accuracy of counting.} Through extensive experiments on sixteen real-world datasets, we empirically show the correctness and convergence of \ours w.r.t the counts of simplets, compared with several baseline methods.
    \item \textbf{Strong power of characterization.} We utilize the counts given by \ours for characterizing real-world SCs and further for SC clustering, showing the strong characterization power of \ours.
\end{itemize}

\smallsection{Reproducibility} {The code and datasets are available at \cite{appendix}.}

\section{Related Work}
\label{sec:related}
\begin{table}[t!]
    \centering
    \caption{Frequently-used notations}
    \label{tab:notation}
    \begin{adjustbox}{max width=0.9\linewidth}
    \begin{tabular}{c|l}
        \toprule
        \textbf{Symbol} & \textbf{Definition} \\
        \midrule
        $\mathcal{K}=(V, E)$ & a simplicial complex with nodes $V$ and edges $E$ \\
        $G_\mathcal{K} = (V_G = V, E_G = E \cap \binom{V}{2})$ & the primal graph of $\mathcal{K}$\\
        $M(\mathcal{K})$ & the set of the maximal simplices in $\mathcal{K}$\\
        $\mathcal{S}^k = \{\mathcal{S}^k_0, \mathcal{S}^k_1, \ldots, \mathcal{S}^k_{s_{k-1}}\}$ & the set of all $s_k$ simplets of size $k$\\
    \bottomrule
    \end{tabular}
    \end{adjustbox}
\end{table}

\smallsection{Simplicial Complex analysis}
In this paper, we focus on a graph representation of SCs~\cite{jonsson2007simplicial},
where 
SCs are defined as higher-order networks with the downward closure property, used to describe higher-order relations in network-like structures~\cite{bianconi2021higher}.
In~\cite{benson2018simplicial}, a triangle-like structure involving three nodes called \textit{simplicial closure} is studied on real-world SCs for higher-order link prediction.
Random walks on SCs are studied in~\cite{schaub2020random} for spectral embedding and estimating the importance of edges.
In~\cite{barbarossa2020topological}, topological signals defined over SCs are studied and applied to wireless network traffic analysis and discrete vector field processing.
Several centrality measures are defined in~\cite{estrada2018centralities} and applied to analyze real-world protein interaction networks.
Also, SCs are used to study systems in
communication~\cite{wang2020social}, epidemic spreading~\cite{li2021contagion}, or social contagion~\cite{iacopini2019simplicial}.
See also~\cite{bianconi2021higher} for a comprehensive introduction to SCs.

\smallsection{Local pattern extraction via (generalized) graphlets}
Extracting local patterns from the abstract graph modelings 
is a common approach to study systems~\cite{cheng2014mining}.
For pairwise graphs, graphlets~\cite{milo2002network, prvzulj2004modeling} describing the interactions among a group of nodes are proposed.
The counts of graphlets are used as characteristic measures of the graph~\cite{prvzulj2007biological,milenkovic2010optimal}, and further used to 
measure graph similarity~\cite{shervashidze2009efficient},
detect anomalies ~\cite{harshaw2016graphprints},
or detect communities~\cite{zhu2021community}.

Recently, graphlet-like patterns are also studied on SCs.
In~\cite{benson2018simplicial}, local patterns in SCs consisting of three interconnected nodes (triangles) are studied.\footnote{Patterns involving four nodes are also briefly discussed in~\cite{benson2018simplicial}.}
A more comprehensive concept called \textit{simplets} generalizing graphlets to SCs is proposed in \cite{preti2022fresco}.
Similar to graphlets, each simplet can be seen as a connected SC without order or node labels, or equivalently an isomorphism class.
There are also several trials on extending graphlets to hypergraphs.
In~\cite{lee2020hypergraph}, connectivity patterns w.r.t the intersections within each group of three edges are studied,
where the patterns 
involve edges as the objects and are limited to groups consisting of three edges only.
In~\cite{lotito2022higher}, another generalization of graphlets on hypergraphs is proposed, where only the patterns consisting of up to four nodes are considered, with hyperedges of sizes larger than four totally ignored.

\begin{figure}[t]
    \centering
    \includegraphics[width=0.868\linewidth]{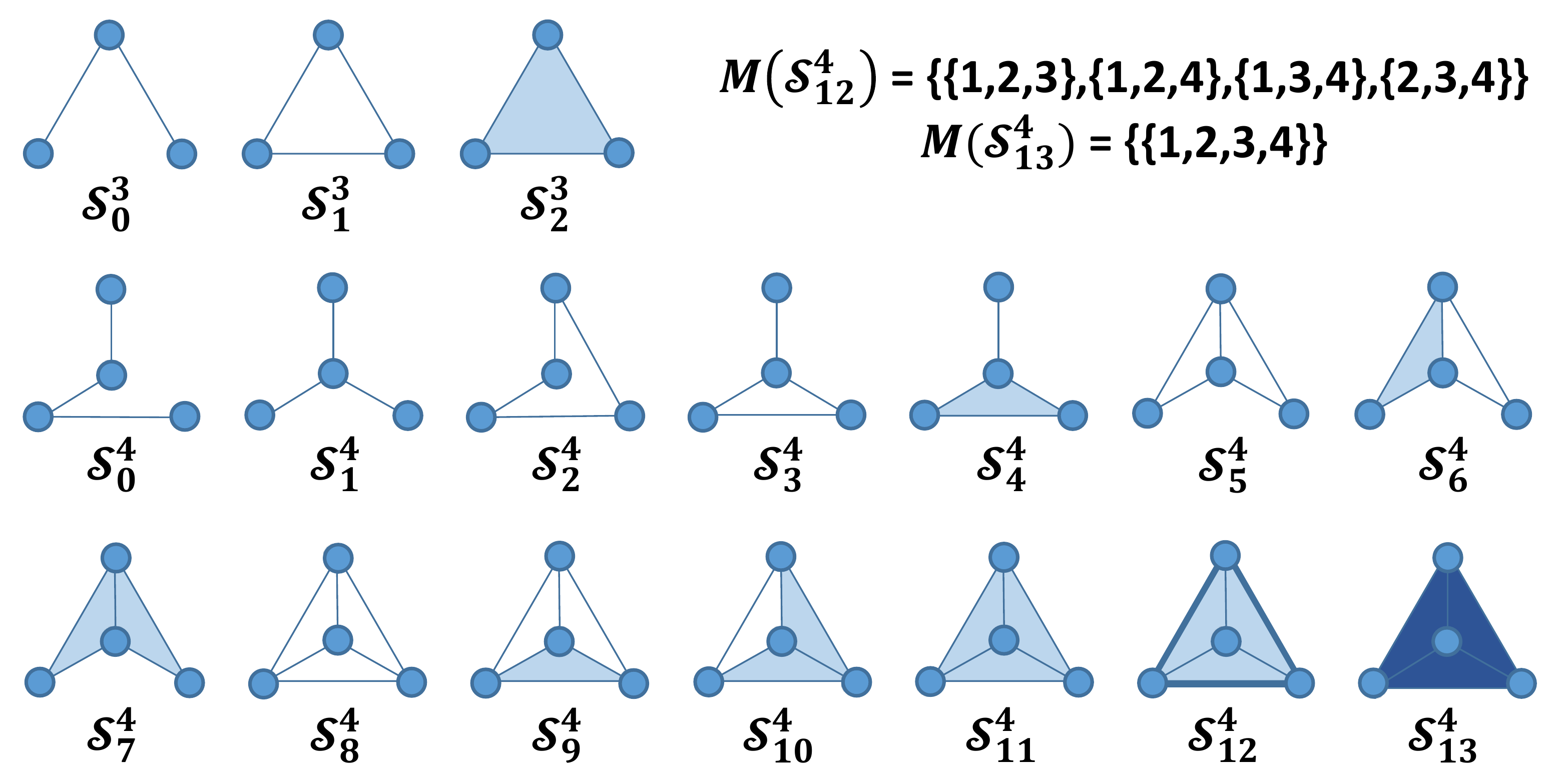}
    \caption{All the simplets of sizes 3 and 4.
    Each light blue triangle shadow represents a group interaction among the three nodes.
    The detailed mathematical descriptions of $\mathcal{S}_{12}^4$ and $\mathcal{S}_{13}^4$ are provided at the top right corner.
    }
    \label{fig:simplet}
\end{figure}

\smallsection{Graphlet and simplet counting}
There are sophisticatedly designed efficient algorithms for exactly counting graphlets of sizes up to five~\cite{ahmed2015efficient,pinar2017escape}.
However, the techniques used in those algorithms are too specific for the graphlets of limited sizes, and cannot be extended for counting simplets.
Recently, for counting graphlets of size more than five,
in~\cite{bressan2017counting, bressan2018motif, bressan2019motivo}, approximate sampling-based methods based on color coding (CC)~\cite{alon1995color} are proposed, showing superiority over methods using Markov chain Monte Carlo (MCMC).

It is even harder to count simplets due to the intrinsically more complicated nature of SCs (compared to pairwise graphs).
In~\cite{preti2022fresco} where the concept of simplets is proposed, the hardness of counting simplets is discussed, and instead of truly counting the occurrences of simplets,
a surrogate measure called \textit{support} is proposed and used for indirect estimation,
and only the problem of computing the surrogate measure is studied.\footnote{The decision version is mainly studied in~\cite{preti2022fresco}, and the problem of exactly computing the supports is discussed in the appendix of~\cite{preti2022fresco}.}
To the best of our knowledge, we are the first to study and propose a practical algorithm for the problem of directly counting the occurrences of simplets.

\section{Concepts and problem statement}
\label{sec:concepts}
In this section, we introduce the 
the concepts used in this paper, 
and present the formal statement of the simplet-counting problem.

\subsection{Concepts}
\smallsection{Basic notations}
Let $\mathbb{N}$ denote the set of positive integers.
Given a set $A$ and $k \in \mathbb{N}$, we use $2^A$ to denote the power set of $A$ 
(i.e., $2^A = \{A': A' \subseteq A\}$),
use $[k]$ to denote $\{0, 1, \ldots, k - 1\}$,
and use $\binom{A}{k}$ to denote the set of all $k$-subsets of $A$ 
(i.e., $\binom{A}{k} = \{A' \subseteq A: |A'| = k\}$).

\smallsection{Hypergraphs}
A \textit{hypergraph} $\mathcal{H} = (V, E)$ consists of 
a node set $V = V(\mathcal{H})$ and
an edge set $E = E(\mathcal{H}) \subseteq 2^V$. %

\smallsection{Simplicial complexes}
A simplicial complex (SC) $\mathcal{K} = (V, E)$ also consists of a node set and an edge set,
while satisfying the \textit{downward closure} property.
That is, for each edge $e \in E$, all the subsets of $e$ are also in $E$
(i.e., $2^e \subseteq E, \forall e \in E$).
Specially, in an SC, each edge $e$ is also called a \textit{simplex},
and 
an induced subcomplex on $V'$ is the SC $\mathcal{K}[V']=(V', E\cap 2^{V'})$.
A simplex $e \in E$ is called a
\textit{maximal} simplex
if there is no strict superset of $e$ in $\mathcal{K}$
(i.e., $\nexists e' \in E$ s.t. $e' \supsetneq e$).
We use $M(\mathcal{K})$ to denote the set of all the maximal simplices in $\mathcal{K}$.
The \textit{primal graph} $G_{\mathcal{K}}$ of $\mathcal{K}$ is a subcomplex of $\mathcal{K}$ consisting of all the simplices of size two (i.e., $G_\mathcal{K} = (V_G = V, E_G = \{e \in E: |e| = 2\} = E \cap \binom{V}{2})$.
An SC is \textit{connected} when its primal graph is connected.

\smallsection{Simplets}
Two SCs $\mathcal{K} = (V, E)$ and
$\mathcal{K}' = (V', E')$ are \textit{isomorphic} (denoted by $\mathcal{K} \simeq \mathcal{K}'$) if there is a bijection $\phi = \phi(\mathcal{K}, \mathcal{K}'): V \rightarrow V'$ such that
$e = (v_1, v_2, \ldots, v_t) \in E$ if and only if
$e' = \phi(e) \in E'$,
where $\phi(e) = (\phi(v_1), \phi(v_2), \ldots, \phi(v_t))$;
we also write $\mathcal{K}' = \phi(\mathcal{K})$,
and $\phi$ is called an \textit{isomorphic bijection} from $\mathcal{K}$ to $\mathcal{K}'$.\footnote{Multiple such bijections may exist, and we let $\phi$ be any of them.}

A \textit{simplet}~\cite{preti2022fresco} 
$\mathcal{S} = (V_{\mathcal{S}}, E_{\mathcal{S}})$ 
of size 
$k = |V_{\mathcal{S}}| \in \mathbb{N}$ 
is a \textit{connected} SC, and
two simplets are seen as the same one if they are isomorphic.
Equivalently, each simplet can be seen as an isomorphic class.
Therefore, WLOG, we assume that for each simplet of size $k$, its node set is $[k]$.
Let $\mathcal{S}^k = \{\mathcal{S}^k_0, \mathcal{S}^k_1, \ldots, \mathcal{S}^k_{s_k-1}\}$ denote the set of all the simplets of size $k$,
where $s_k = |\mathcal{S}^k|$ is the number of such simplets, for each $k \in \mathbb{N}$.
Under the isomorphic equivalence relation, 
$s_k = 1, 1, 3, 14, 157, 15942$ 
for $k = 1, 2, 3, 4, 5, 6$, respectively.\footnote{See
{Appendix~\ref{sec:alg_A}}
for an algorithm generating all the simplets of size $k$ based on all the graphlets of the same size.}
In Figure~\ref{fig:simplet}, we list all simplets of size $3$ and $4$.
A \textit{graphlet}~\cite{bressan2018motif, prvzulj2004modeling} $\mathcal{L} = (V_{\mathcal{L}}, E_{\mathcal{L}})$ can be seen as a special simplet that is a pairwise graph (i.e., $|e| = 2, \forall e \in E_{\mathcal{L}}$).

Given an SC $\mathcal{K} = (V_\mathcal{K}, E_\mathcal{K})$ 
and a simplet $\mathcal{S}$ of size $k$, we say there is an \textit{occurrence} of $\mathcal{S}$ on $X \in \binom{V}{k}$ in $\mathcal{K}$ 
(the event is denoted by $OC(\mathcal{S}, X; \mathcal{K})$), if 
the \textit{induced} subcomplex $\mathcal{K}[X]$ is isomorphic to $\mathcal{S}$.\footnote{{In this paper, only induced subcomplexes are counted following the typical definition for graphlets~\cite{hovcevar2014combinatorial}, while non-induced ones are also counted in~\cite{preti2022fresco}. The two kinds of counts are equivalent while non-induced counting includes unnecessary repetition.}}
The total \textit{number of occurrences} of $\mathcal{S}$ in $\mathcal{K}$ is
\begin{align*}\label{eq:total_number_of_occ}
N_{oc}(\mathcal{S}; \mathcal{K}) =
\left|\left\{X \in  \binom{V}{k}: OC(\mathcal{S}, X; \mathcal{K})\right\}\right|.
\end{align*}

\smallsection{Treelets and colorful treelets}
Given an SC $\mathcal{K}=(V, E)$ and $k \in \mathbb{N}$,
we apply a $k$-coloring $f: V \rightarrow [k]$,
where to each $v \in V$ a color $f(v) \in [k]$ is assigned.
We define \textit{treelets} as special cases of simplets that are trees.
We use $\mathcal{T}^k$ to denote the set of all the size-$k$ treelets.
We say a treelet $\mathcal{T} \in \mathcal{T}^k$ is a \textit{colorful} treelet~\cite{bressan2017counting} if all the $k$ colors are used on the $k$ nodes in $\mathcal{T}$ (i.e., $\{f(v): v \in V_{\mathcal{T}}\} = [k]$).

The frequently-used notations are summarized in Table~\ref{tab:notation}.
In the notations, the input SC $\mathcal{K}$ can be omitted when the context is clear.

\begin{algorithm}[t]
    \caption{Brute-force enumeration of simplets}
    \label{algo:naive_enumeration}
    \SetKwInput{KwInput}{Input}
    \SetKwInput{KwOutput}{Output}
    
    \KwInput{(1) $\mathcal{K} = (V, E)$: the input simplicial complex; \\
    \quad\quad\quad (2) $k$: the considered size of simplets; \\
    \quad\quad\quad (3) $\mathcal{S}^k$: the set of all the simplets of size $k$}
    \KwOutput{$N_{oc}(\mathcal{S}^k_{i}; \mathcal{K}), \forall i \in [s_k]$: the count of each simplet}
    \SetKwProg{Fn}{Function}{:}{\KwRet}
    
    $N_{oc}(\mathcal{S}^k_{i}; \mathcal{K}) \leftarrow 0, \forall i \in [s_k]$ \Comment*[f]{Initialization} \\
    \ForEach{$V_k \in \binom{V}{k}$}{ \label{algo:naive_enumeration:all_k_subsets}
        \ForEach{bijection $\phi: V_k \rightarrow [k]$ \label{algo:naive_enumeration:all_bijections}}{
            \If{$\phi(\mathcal{K}[V_k]) \in \mathcal{S}^k$}{ \label{algo:naive_enumeration:in_Sk_check}
                $N_{oc}(\phi(\mathcal{K}[V_k]); \mathcal{K}) \leftarrow N_{oc}(\phi(\mathcal{K}[V_k]); \mathcal{K}) + 1$ \label{algo:naive_enumeration:increment} \\
            }
        }
    }
    \Return{$N_{oc}(\mathcal{S}^k_{i}; \mathcal{K}), \forall i \in [s_k]$} \\
\end{algorithm}

\subsection{Problem Statement}\label{sec:preliminaries:problem}
We are now ready to present the formal statement of the problem that we study in this paper, where we aim to count the occurrences of each simplet of a given size $k$ in a given SC $\mathcal{K}$.

\begin{problem}\label{problem:simplet_counting}
    Given an SC $\mathcal{K} = (V, E)$ and $k \in \mathbb{N}$, 
    we aim to count the number $N_{oc}(\mathcal{S}^k_{i}; \mathcal{K})$ of occurrences of $\mathcal{S}^k_{i}$ in $\mathcal{K}$, for each $i \in [s_k]$.
\end{problem}

Problem~\ref{problem:simplet_counting} extends the well-known and widely-studied counterpart problem of counting graphlets~\cite{marcus2012rage, wernicke2006fanmod, hovcevar2014combinatorial, ahmed2015efficient,rahman2014graft,bhuiyan2012guise, han2016waddling, slota2013fast, wang2014efficiently, bressan2018motif,bressan2019motivo}.
Due to the quadratic nature of graphlets (essentially, of pairwise relations), and the exponential nature of simplets (essentially, of group relations), Problem~\ref{problem:simplet_counting} is intrinsically more difficult than the counterpart on graphlets, while inheriting the \#W[1]-hardness~\cite{jerrum2015parameterised}.
Specifically, the brute-force enumeration for graphlet counting takes $O(k^2 n^k)$ time~\cite{bressan2018motif};
while for simplet counting it takes $O(2^k n^k)$ time, where $k$ is the given size of considered graphlets or simplets and $n$ is the number of nodes in the input graph or SC.
We present the enumeration process for simplet counting in Algorithm~\ref{algo:naive_enumeration},
where
we first enumerate all the $k$-subsets $V_k$ of the node set (Line~\ref{algo:naive_enumeration:all_k_subsets}) and 
all the bijections from $V_k$ to $[k]$ (Line~\ref{algo:naive_enumeration:all_bijections})
to check which simplet the induced subcomplex on $V_k$ corresponds to (this takes $O(2^k)$ time since a simplet of size $k$ has $O(2^k)$ edges) and increment the counting accordingly (Lines~\ref{algo:naive_enumeration:in_Sk_check} and \ref{algo:naive_enumeration:increment}),
which gives the total time complexity $\binom{|V|}{k} \cdot k! \cdot 2^k = O(2^k |V|^k)$.\footnote{See~\cite{preti2022fresco} for more discussion on the hardness of this problem.}

Due to the prohibitive time complexity of the brute-force method, it is imperative to have a faster method for simplet counting.
Unfortunately, the existing techniques for exact graphlet counting rely on sophisticated designs tailored for specific problems and cannot be directly extended to simplet counting.
Therefore, instead of exact counting,
we aim to propose an approximate algorithm with high accuracy.

\section{Proposed method}
\label{sec:method}
In this section, we introduce in detail our algorithm \textbf{\ours} (\OURS).
We present the algorithmic details based on color coding (CC)~\cite{alon1995color} for counting the occurrences of each specific simplet of size $k$.
The proposed algorithm \ours consists of four phases: building phase, sampling phase, scanning phase, and lastly, matching phase,
which extends the CC-based graphlet-counting algorithms~\cite{bressan2017counting, bressan2018motif, bressan2019motivo} to simplet counting with essential technical novelty, especially in the scanning and matching phases.
The theoretical analysis is provided in Section~\ref{sec:math}, where we show the unbiasedness, convergence, and time/space complexity of \ours.

\subsection{Overview of \ours}

\ours is composed of four steps: building, sampling, scanning, and matching. For a given SC $\mathcal{K}$ and size $k$, we (a) sample a (colorful and non-induced) \textit{tree} $T=(V_T,E_T)$ such that $|V_T|=k$, uniformly at random (sampling step), (b) find an induced subcomplex $\mathcal{K}[V_T]$ based on the node set of $T$ (scanning step), (c) match it to an isomorphic simplet $\mathcal{S}\in \mathcal{S}_k$ (matching step), and (d) repeat this procedure. To prepare such a sampling step, the information of trees is needed, which is stored in a table (building step). The entire process is presented on Algorithm~\ref{algo:total} after explaining the details of each step. See Appendix~\ref{app:example} for a toy example.

\subsection{Building and sampling colorful treelets}

For the first two steps (building and sampling) the key techniques are from color coding~\cite{alon1995color};
based on the corresponding steps of the graphlet counting algorithm in~\cite{bressan2017counting, bressan2018motif, bressan2019motivo}, we adapt them for simplet counting.
We describe the details of the two steps below.

\smallsection{Building}
A primal graph $G_\mathcal{K}$ of an SC $\mathcal{K}$ is given as an input of the building step 
together with the considered size $k$ of simplets.
For each tuple $(v, \mathcal{T}, S)$ consisting of a node $v \in V$, a treelet $\mathcal{T}$ of size at most $k$, and a set of colors $S$ of the same size as $\mathcal{T}$ (thus $\mathcal{T}$ colored by $S$ is colorful), we record the number $C(v, \mathcal{T}, S)$ of occurrences of $\mathcal{T}$ colored by $S$ rooted at $v$;
at the same time, we count the total number $N_{ct}$ of occurrences of the colorful size-$k$ treelets.

Specifically, we first apply a $k$-coloring to the input $G_\mathcal{K}$ by coloring each node with a color in $[k]$ uniformly at random.
For counting the occurrences of color treelets, the key idea is recursively computing each $C(v, \mathcal{T}, S)$ from each pair of $C(v, \mathcal{T}_1, S_1)$ and $C(v, \mathcal{T}_2, S_2)$ with $\mathcal{T} = \mathcal{T}_1 \sqcup \mathcal{T}_2$ and $S = S_1 \sqcup S_2$ until the size of $\mathcal{T}$ reaches $k$, where $\sqcup$ denotes the operation of disjoint union.

\smallsection{Sampling}
Besides the input primal graph $G_\mathcal{K}$ and the size $k$, the output of the building step (the numbers of occurrences of rooted colorful treelets $C(\cdot, \cdot, \cdot)$ and the total number $N_{ct}$ of occurrences of the colorful treelets of size $k$) as well as a user-defined number $x$ of samples to draw are given.
We sample a set $T_{ct}$ of $x$ occurrences of colorful treelets uniformly at random among all the $N_{ct}$ occurrences of colorful treelets of size $k$, where each occurrence is output as a set of nodes.
The key idea is that each occurrence of color treelet is sampled with a probability proportional to $C(v, \mathcal{T}, [k])$, which is achieved by recursively sampling subtrees.\footnote{The sampled node sets might be duplicated, and then the output $T_{ct}$ is a multiset.}

\begin{lemma}\label{lem:build_sample}
    Given $G_\mathcal{K} = (V_G, E_G)$, $k \in \mathbb{N}$, and a user-defined number $x$ of samples,
    the building and sampling steps (Algorithms~\ref{algo:build} and \ref{algo:sample}) 
    output $T_{ct}$ consisting of $x$ node sets $V_k \in V_{con}$
    in $O(c^k|E_G| + xk)$ time and 
    $O(c^k|E_G|)$ space 
    for some absolute constant $c > 0$,
    where $V_{con} = \{V_k \in \binom{V_G}{k}: G_\mathcal{K}[V_k] \text{~is connected}\}$.
    Moreover, in each iteration of sampling, a $V_k$ is sampled with a probability proportional to
    $n_{st}(G_\mathcal{K}[V_k])$, the number of spanning trees of $G_\mathcal{K}[V_k]$.
\end{lemma}
\begin{proof}
    Refer to Appendix~\ref{sec:pfs_A} for all the proofs.
\end{proof}

As mentioned above, the building and sampling phases are mainly based on~\cite{bressan2017counting, bressan2018motif, bressan2019motivo} for graphlet counting, and extended by us for simplet counting.
This extension works because color coding samples among all connected induced subgraphs, and
a subcomplex of an SC induced on a node set is connected if and only if 
the induced subgraph of the primal graph of the SC on the same node set is connected.
For completeness, we provide the detailed processes in Algorithms~\ref{algo:build} and \ref{algo:sample} in Appendix~\ref{sec:alg_A}.
See~\cite{bressan2017counting, bressan2018motif, bressan2019motivo} for more details.

\begin{algorithm}[t]
    \small
    \caption{SC3-scan}\label{algo:scan}
    \SetKwInput{KwInput}{Input}
    \SetKwInput{KwOutput}{Output}
    \KwInput{(1) $T_{ct}$: the sampled occurrences of colorful treelets\\
    \quad\quad\quad (2) $\mathcal{K}=(V, E)$: the input SC
    \label{algo:scan:input}}
    \KwOutput{$M(\mathcal{K}[V_T]), \forall V_T \in T_{ct}$: the set of maximal simplices in the subcomplex of $\mathcal{K}$ induced on each sampled $V_T$}
    
    $M(\mathcal{K}[V_T]) \leftarrow \emptyset, \forall V_T \in T_{ct}$ \label{algo:scan:init} \Comment*[f]{Initialization} \\
    \ForEach{$V_T \in T_{ct}$ \label{algo:scan:for_each_sample} }{
        \ForEach{$\sigma \in M(\mathcal{K})$ s.t. $|V_T \cap \sigma| > 1$ \label{algo:scan:for_each_sigma}} { 
            $k_\sigma \leftarrow V_T \cap \sigma$ \\
            \If{$\nexists q \in M(\mathcal{K}[V_T])$ s.t. $k_\sigma \subseteq q$ \label{algo:scan:if_maximal}}{
                $M(\mathcal{K}[V_T]) \leftarrow M(\mathcal{K}[V_T]) \cup \{k_\sigma\} \setminus \{q \in M(\mathcal{K}[V_T]) : q \subsetneq k_\sigma\}$ \label{algo:scan:maximal} \\
            }
        }
    }
    \Return{$M(\mathcal{K}[V_T]), \forall V_T \in T_{ct}$} \label{algo:scan:return}
\end{algorithm}

\subsection{Scanning the maximal simplices}
The building and sampling steps give us a set $T_{ct}$ of occurrences of colorful treelets.
In the scanning step, we aim to find for each occurrence $V_T \in T_{ct}$ the set of maximal simplices in the induced subcomplex of $\mathcal{K}$ on $V_T$, where $\mathcal{K}$ is the input SC.

In Algorithm~\ref{algo:scan}, we provide the detailed process of scanning.
We first initialize the set $M(\mathcal{K}[V_T])$ of maximal simplices as empty for each sampled $V_T$ (Line~\ref{algo:scan:init}).
Then for each $V_T$ (Line~\ref{algo:scan:for_each_sample}), 
and for each maximal simplex $\sigma$ intersecting with $V_T$ with more than one node (Line~\ref{algo:scan:for_each_sigma}) 
that is maximal in the current $M(\mathcal{K}[V_T])$ (Line~\ref{algo:scan:if_maximal}),
we add the intersection $k_\sigma = V_T \cap \sigma$ into $M(\mathcal{K}[V_T])$ while removing all the strict subsets of $k_\sigma$ for computational and memory efficiency without affecting correctness (Line~\ref{algo:scan:maximal}).
Note that we utilize the fact that $M(\mathcal{K}[V_T]) \subset \{V_T \cap \sigma: \sigma \in M(\mathcal{K})\}$ to avoid checking all simplices.
Also, note that we store only $M(\mathcal{K})$ for each SC $\mathcal{K}$.

\begin{lemma}\label{lem:scan}
    Given $T_{ct}$ and $\mathcal{K} = (V, E)$,
    Algorithm~\ref{algo:scan} correctly outputs $M(\mathcal{K}[V_T])$ for all $V_T \in T_{ct}$ 
    in $O(|M(\mathcal{K})|\hat{M}_{ct})$ time and $O(\hat{M}_{ct})$ space,
    where $\hat{M}_{ct} = \sum_{V_T \in T_{ct}} |M(\mathcal{K}[V_T])|$.
\end{lemma}

\begin{algorithm}[t]
    \small
    \caption{SC3-match}\label{algo:match}
    \SetKwInput{KwInput}{Input}
    \SetKwInput{KwOutput}{Output}
    \KwInput{(1) $T_{ct}$: the sampled occurrences of colorful treelets \\
    \quad\quad\quad (2) $M(\mathcal{K}[V_T]), \forall V_T \in T_{ct}$: the set of maximal simplices \\
    \quad\quad\quad\quad\ \  in the subcomplex induced on each sampled $V_T$ \\
    \quad\quad\quad (3) $k$: the considered size of simplets/treelets \\
    \quad\quad\quad (4) $\mathcal{S}^k$: the set of all the simplets \\
    \quad\quad\quad (5) $N_{ct}$: the total count of the colorful treelets}
    \KwOutput{$\tilde{N}_{oc}(\mathcal{S}), \forall \mathcal{S} \in \mathcal{S}^k$: the estimated count of each simplet}
    $\tilde{N}_{oc}(\mathcal{S}) \leftarrow 0, \forall \mathcal{S} \in \mathcal{S}^k$ \label{algo:match:init} \Comment*[f]{Initialization} \\ 
    $f(\phi(M(\mathcal{S}))) \leftarrow \mathcal{S}, \forall \mathcal{S} \in \mathcal{S}^k, \forall \text{bijection}~\phi: [k] \rightarrow [k]$ \label{algo:match:perm_eq_class} \\
    \ForEach{$V_T \in T_{ct}$ \label{algo:match:for_each_sample}}{
        $\mathcal{S}_T \leftarrow f(M(\mathcal{K}[V_T]))$ \label{algo:match:perm_map_recover} \\
        $\tilde{N}_{oc}(\mathcal{S}_T) \leftarrow \tilde{N}_{oc}(\mathcal{S}_T) + \frac{1}{{n_{st}(G_{\mathcal{S}_T})}} \cdot \frac{N_{ct}}{|T_{ct}|} \cdot \frac{k^k}{k!}$ \label{algo:match:count_inc} \\
    }
 \Return $\tilde{N}_{oc}(\mathcal{S}), \forall \mathcal{S} \in \mathcal{S}^k$
 \label{algo:match:end}
\end{algorithm}

\subsection{Matching the simplets}
After the scanning process, for each sampled occurrence $V_T$ of colorful treelet, we have the set $M(\mathcal{K}[V_T])$ of maximal simplices in the subcomplex induced on $V_T$.
We now aim to match each $M(\mathcal{K}[V_T])$ to a simplet $\mathcal{S} \in \mathcal{S}^k$.

Finding the isomorphic simplets is done by precomputing all possible cases generated from permutation (Line~\ref{algo:match:perm_eq_class}).
For each sampled $V_T \in T_{ct}$, we recover the whole SC from its maximal simplices and find the corresponding simplet
(Line~\ref{algo:match:perm_map_recover}),
and increase the estimated count by a normalized value ($\frac{1}{{n_{st}(G_{\mathcal{S}_T})}} \cdot \frac{N_{ct}}{|T_{ct}|} \cdot \frac{k^k}{k!}$ on Line~\ref{algo:match:count_inc}), where $n_{st}(G_{\mathcal{S}_T})$ is the number of spanning trees precomputed by e.g., \cite{tutte2001graph} with $O(c^k)$ time where $c>0$ is a constant. 
The term $n_{st}(G_{\mathcal{S}_T})$ is used because each $V_T$ can be sampled from $n_{st}(G_{\mathcal{S}_T})$ different trees in the sampling step,
the term $\frac{N_{ct}}{|T_{ct}|}$ is the proportion of the sampled occurrences,
and the term $\frac{k^k}{k!}$ comes from the fact that for $k$-set of nodes, there are $k!$ ways of $k$-coloring for it to be colorful while there are $k^k$ ways in total.

\begin{lemma}\label{lem:match}
    Given $T_{ct}$, $M(\mathcal{K}[V_T]), \forall V_T \in T_{ct}$, $k$, $S_k$, and $N_{ct}$,
    Algorithm~\ref{algo:match} takes
    $O(k!\hat{M}_k + |T_{ct}|)$ time and $O(\hat{M}_k + \hat{M}_{ct})$ space,
    where $\hat{M}_k = \sum_{\mathcal{S} \in \mathcal{S}^k} |M(\mathcal{S})|$ and
    $\hat{M}_{ct} = \sum_{V_T \in T_{ct}} |M(\mathcal{K}[V_T])|$. Here, $\hat{M}_k$ is a function of $k$: $\hat{M}_4=47$ and $\hat{M}_5=807$.
\end{lemma}

\subsection{Theoretical Analysis} \label{4.3}\label{sec:math}

Now we conclude the whole process of \ours in Algorithm~\ref{algo:total}, and theoretically analyze its properties.
Specifically, we shall show the 
\begin{itemize}[leftmargin=*]
    \item \textbf{Unbiasedness:} the output $\tilde{N}_{oc}(\mathcal{S})$ is an unbiased estimator of the ground truth $N_{oc}(\mathcal{S})$, for each $\mathcal{S}$;
    \item \textbf{Convergence:} the output $\tilde{N}_{oc}(\mathcal{S})$ converges to the ground truth $N_{oc}(\mathcal{S})$, for each $\mathcal{S}$, as the number of repeated trials increases;
    notably, for a single trial of \ours (a fixed coloring), when the number $x$ of samples increases, the output converges. 
    \item \textbf{Complexities:} the time and space complexities of \ours.
\end{itemize}

\begin{theorem}[unbiasedness]\label{thm:unbiasedness}
    Given $k$, $\mathcal{K}$, $\mathcal{S}^k$, and any $x$,
    for each $\mathcal{S} \in \mathcal{S}^k$,
    the $\tilde{N}_{oc}(\mathcal{S})$ given by Algorithm~\ref{algo:total}
    satisfies that
    $\mathbb{E}[\tilde{N}_{oc}(\mathcal{S})] = N_{oc}(\mathcal{S})$.
\end{theorem}

\begin{theorem}[convergence] \label{thm:convergence}
    Given any $k$, $\mathcal{K}$, and $\mathcal{S}^k$,
    for each $\mathcal{S} \in \mathcal{S}^k$,
    let $\tilde{N}_{oc}^i(\mathcal{S})$ denote the output by Algorithm~\ref{algo:total} in the $i$-th trial.
    For any $\epsilon, \lambda > 0$, there exists $R_t > 0$ such that
    if $R > R_t$, then
    $\Pr[|\sum_{i \in [R]} \tilde{N}_{oc}^i(\mathcal{S})/R - N_{oc}(\mathcal{S})| \leq \lambda] \geq 1 - \epsilon$, for any $x \geq 1$.
    For a single trial, $\Pr[|\tilde{N}_{oc}(\mathcal{S}) - N_{oc}(\mathcal{S}) r_{color}| \leq \lambda\sigma ]\geq 1-\frac{1}{\lambda^2}$
    where $\sigma^2=Var[\tilde{N}_{oc}(\mathcal{S})]=O(\frac{1}{x})$ (in terms of $x$ only), and  $r_{color}$ is the ratio between the actual count of occurrences of colorful treelets corresponding to $\mathcal{S}$ and the expected count. 
\end{theorem}

\begin{theorem}[complexities] \label{thm:complexities}
    Given $k$, $\mathcal{K}=(V, E)$, $\mathcal{S}^k$, and $x$,
    Algorithms~\ref{algo:total} takes 
    $O(c^k |E_G| + x |M(\mathcal{K})|^2 + k! \hat{M}_k)$ time and 
    $O(c^k |E_G| + x |M(\mathcal{K})| + \hat{M}_k)$ space for some absolute constant $c > 0$,
    where $E_G = E \cap \binom{V}{2}$ and $\hat{M}_k=\sum_{\mathcal{S}\in \mathcal{S}^k} |M(\mathcal{S})|.$
\end{theorem}

For Problem~\ref{problem:simplet_counting}, high complexity w.r.t. $k$ is inevitable since $|\mathcal{S}_k|$ increases exponentially w.r.t. $k$. However, \ours is scalable w.r.t factors other than $k$ and empirically much faster than the competitors.

\begin{algorithm}[t]
    \caption{\textbf{\ours}: \OURS}
    \label{algo:total}
    \SetKwInput{KwInput}{Input}
    \SetKwInput{KwOutput}{Output}
    
    \KwInput{(1) $k$: the considered size of graphlets and simplets\\
    \quad\quad\quad (2) $\mathcal{K}$: the input SC \\
    \quad\quad\quad (3) $\mathcal{S}^k$: the set of all the simplets \\
    \quad\quad\quad (4) $x$: the user-defined number of samples}
    \KwOutput{$\tilde{N}_{oc}(\mathcal{S}), \forall \mathcal{S} \in \mathcal{S}^k$: the estimated count of each simplet}
    
    \SetKwProg{Fn}{Function}{:}{\KwRet}
    $C, N_{ct} \leftarrow$ Alg.~\ref{algo:build} with $k$ and $G_\mathcal{K}$ \label{algo:total:build} \Comment*[f]{Building} \\ 
    $T_{ct} \leftarrow$ Alg.~\ref{algo:sample} with $k$, $G_\mathcal{K}$, $C$, $N_{ct}$ and $x$ \label{algo:total:sample} \Comment*[f]{Sampling} \\ 
    $M(\mathcal{K}[V_T]) \leftarrow$ Alg.~\ref{algo:scan}  with $T_{ct}$ and $\mathcal{K}, \forall V_T \in T_{ct}$ \label{algo:total:scan} \Comment*[f]{Scanning} \\ 
    $\tilde{N}_{oc}(\mathcal{S}) \leftarrow$ Alg.~\ref{algo:match} with $T_{ct}$, $M(\mathcal{K}[V_T]), \forall V_T \in T_{ct}$, $k$, $\mathcal{S}^k$, and $N_{ct}, \forall \mathcal{S} \in \mathcal{S}^k$ \label{algo:total:match} \Comment*[f]{Matching} \\ 
    \Return{$\tilde{N}_{oc}(\mathcal{S}), \forall \mathcal{S} \in \mathcal{S}^k$} \label{algo:total:return}
\end{algorithm}

\section{Characterization of simplicial complex using simplet counts}
\label{sec:characterization}
In this section, we show (1) the counts given by \ours are accurate in that they are close to the ground truth counts, and (2) the counts given by \ours can be used for characterizing SCs.

\smallsection{Characteristic Profile (CP)}
We use a measure called \textit{characteristic profile}~\cite{milo2004superfamilies} (CP).
which measures
the significance of each simplet $\mathcal{S}_i^k \in \mathcal{S}^k$ ($i \in [s_k]$),
for a given $k \in \mathbb{N}$ in a given SC $\mathcal{K}$.
We first normalize the count to get the \textit{ratio}
$\tilde{n}_{oc}(\mathcal{S}_i^k; \mathcal{K}) =
\tilde{N}_{oc}(\mathcal{S}_i^k; \mathcal{K}) / \sum_{\mathcal{S}' \in \mathcal{S}^k} \tilde{N}_{oc}(\mathcal{S}'; \mathcal{K})$ of each $i \in [s_k]$.
Then we define the \textit{significance vector} $\mu = \mu(\mathcal{K}) = (\mu_0, \mu_1, \ldots, \mu_{s_k-1})$ of $\mathcal{K}$ by
\begin{equation}
    \mu_{i} = \frac{\tilde{n}_{oc}(\mathcal{S}_i^k;\mathcal{K})-\tilde{n}_{oc}(\mathcal{S}_i^k;\mathcal{K}_\mathcal{R})}{\tilde{n}_{oc}(\mathcal{S}_i^k;\mathcal{K})+\tilde{n}_{oc}(\mathcal{S}_i^k;\mathcal{K}_\mathcal{R})+\epsilon},
\end{equation}
where $\mathcal{K}_\mathcal{R}$ is any random SC generated by a null model (we will define the null model that we use later) from $\mathcal{K}$, and $\epsilon > 0$ is a small enough constant.
In our experiments, we set $\epsilon = 10^{-3}$.
Based on the significance vector, we compute the CP~\cite{milo2004superfamilies} of $\mathcal{K}$ as a normalized significance vector:
\begin{equation*}
    CP_{i} = \frac{\mu_{i}}{\sqrt{\sum_{i \in [s_k]} {\mu_{i}}^2}}.
\end{equation*}
The CP of an SC contains the information of local patterns and allows us to compare multiple SCs.

\smallsection{Null model}
As mentioned above, a null model is required to compute the significance vector and CP of an SC.
Regarding the choice of the null model, 
we aim to preserve the number of simplices and the size of each simplex.
Given an SC $\mathcal{K} = (V, E)$, we extract all its maximal simplices.
First, we decide a size $t$, where each size $t$ is chosen with a probability 
$\frac{|\{e \in E:|e|= t\}|}{|E|}$;
then, we choose a pair of maximal simplices of size $t$ uniformly at random among all the pairs of size-$t$ maximal simplices;
finally, we repeatedly switch nodes independently between the pair for $\lfloor i/2 \rfloor$ times.
We obtain all the maximal simplices of the random complex $\mathcal{K}_\mathcal{R}$ by repeating the above procedure $c_{shuffle}|E|$ times and expanding all the maximal simplices.
We use $c_{shuffle} = 1,000$ in our experiments.

\section{Experiments}
\label{sec:experiments}
\begin{figure}[t]
    \vspace{-2.5mm}
    \centering
    \includegraphics[width=0.95\linewidth]{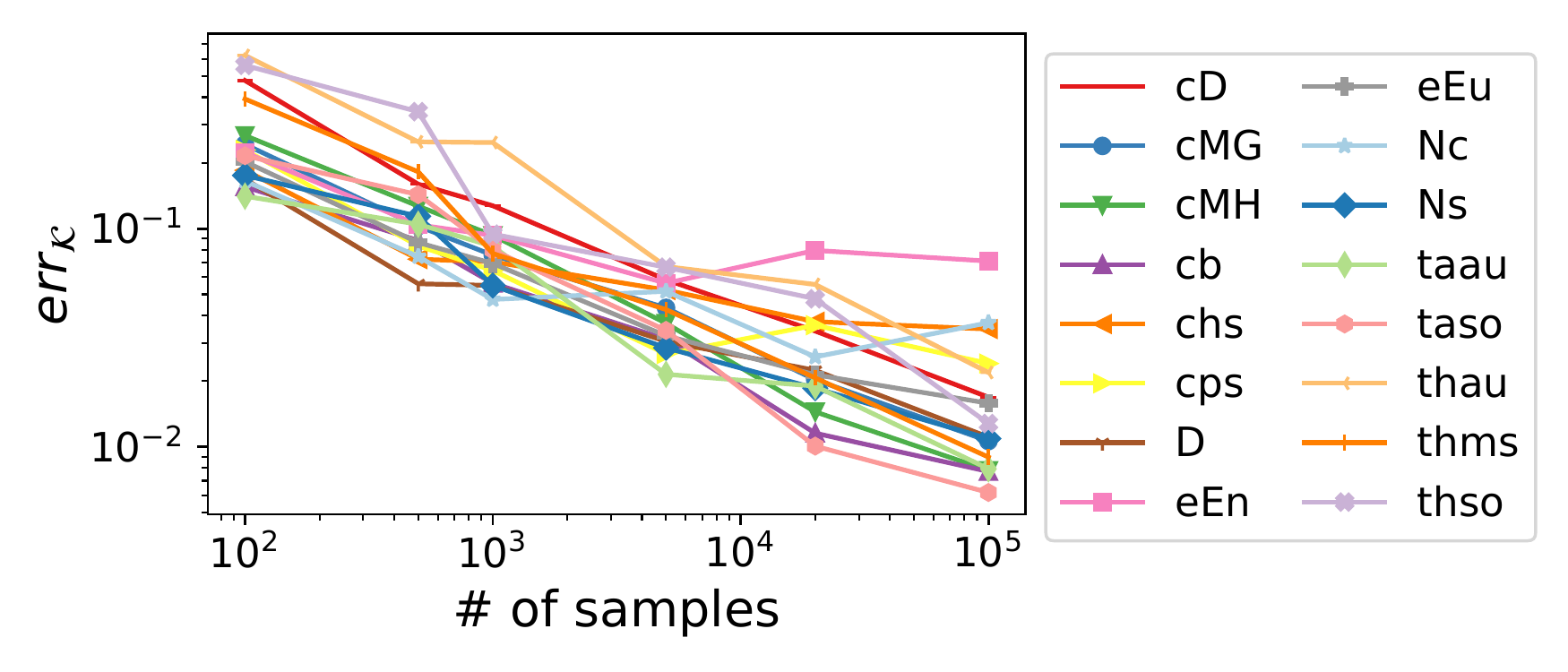} \\
    \caption{\underline{\smash{\ours with $x = 100,000$ shows $err_\mathcal{K}$ lower than $0.05$,}} \underline{\smash{except for the \emailEnron dataset.}} We report the mean value of the errors over 5 trials on each dataset.}
    \label{fig:mae}
\end{figure}

We performed experiments on sixteen real-world SCs using \ours and several baseline methods, aiming to answer the following questions:
\begin{itemize}[leftmargin=*]
    \item \textbf{Q1. Accuracy:} How accurate are the counts of simplets obtained by \ours?
    How well do the counts converge to the ground truth values, as the number of samples increases?
    \item \textbf{Q2. Scalability and speed:} How fast is \ours compared to the baseline algorithms?
    How does the running time of \ours grow as the number of samples increases? 
    \item \textbf{Q3. Characterization power across domains:} How well does the characteristic profile obtained from the counts by \ours cluster the real-world SCs from different domains?
\end{itemize}

\subsection{Experimental Setting}

\smallsection{Machine} We performed all the experiments on a machine with a 3.7GHz Intel i5-9600K CPU and 64GB memory. 

\smallsection{Dataset} We used $16$ real-world  simplicial-complex datasets.
We provide the basic statistics of them in Appendix~
\ref{sec:data}.

\smallsection{Competitors} We compared \ours with two existing algorithms:  designed for SCs:
(1) \Benson~\cite{benson2018simplicial} and
(2) \FreSCo~\cite{preti2022fresco}. %

\begin{itemize}[leftmargin=*]
    \item \textbf{\Benson} exactly counts 3- or 4-node configurations using combinatorial methods.
    Notably, each simplet may correspond to zero, one, or multiple node configurations of the same size.
    See Appendix~\ref{app:base} for the detailed correspondent relations between simplets and node configurations.
    \item \textbf{\FreSCo} indirectly estimate the count of each simplet using a surrogate measure called \textit{support} (see Appendix~\ref{app:base} or~\cite{preti2022fresco} for the formal definition).
    The support satisfies that if two simplices $\sigma_1 \subseteq \sigma_2$ then the support of $\sigma_1$ is at least that of $\sigma_2$.
    We compare \ours with 
    \FreSCo that exactly computes the support of each simplet.
    See Appendix~\ref{app:base} (esp. Table~\ref{tab:fresco}) for comparisons between the exact or estimated counts and the supports from \FreSCo, where we observe that the supports computed in $10$ hours (the time limit that we set) are not strongly related to the exact counts.
\end{itemize}

\smallsection{Implementations}
We implemented \ours in C++. It supports multi-threading for the building and scanning steps, and we set the number of threads to 6.
For \Benson, we used the open-source implementation in Julia provided by the authors.
For \FreSCo, we used the open-source implementation in Java provided by the authors.

\begin{figure}[t]
    \vspace{-3mm}
    \centering
    \includegraphics[width=0.9\linewidth]{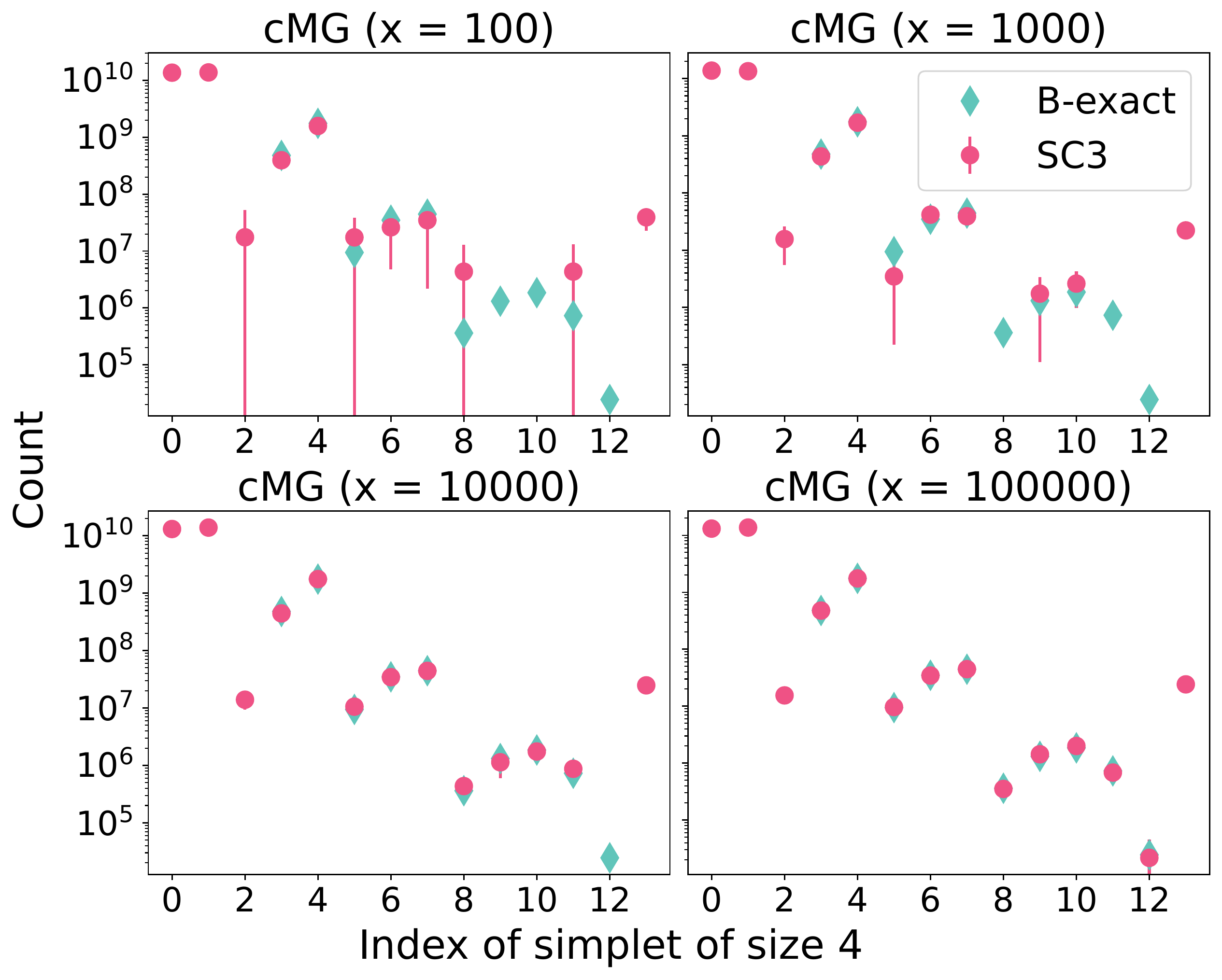} \\
    \caption{\underline{\smash{The estimated counts by \ours are closer to the exact}} 
    \underline{\smash{counts as the number of samples increases.}} For $k = 4$, we report the mean value of each the counts of each simplet as a point, and the standard deviation as an error bar.}
    \label{fig:conv4}
\end{figure}

\subsection{\textbf{Q1. Accuracy}}

To evaluate the accuracy, we compared the ground truth counts computed by \Benson and the estimated counts obtained by \ours while varying the number of samples from $100$ to $100,000$.
For the error measure, we used a normalized error $err_{\mathcal{K}}$ defined as
\begin{align*}
 err_{\mathcal{K}}=\frac{\sum_{i \in [s_k]} |N_{oc}(\mathcal{S}_i^k; \mathcal{K})-\tilde{N}_{oc}(\mathcal{S}_i^k; \mathcal{K})|}{\sum_{i \in [s_k]}N_{oc}(\mathcal{S}_i^k; \mathcal{K})}.
\end{align*}
Since \Benson only provides the exact counts when $k \leq 4$, we fixed $k = 4$.
As seen in Figure~\ref{fig:mae}, the count of each $\mathcal{S}_i^k \in \mathcal{S}^k$ converges to the exact count, as the number of samples increases. 
Specifically, on $15$ out of $16$ datasets, the normalized error $err_{\mathcal{K}}$ is below $0.05$, showing the high accuracy of \ours on real-world datasets.

\begin{figure}[t]
    \vspace{-2mm}
    \centering
    \includegraphics[width=0.88\linewidth]{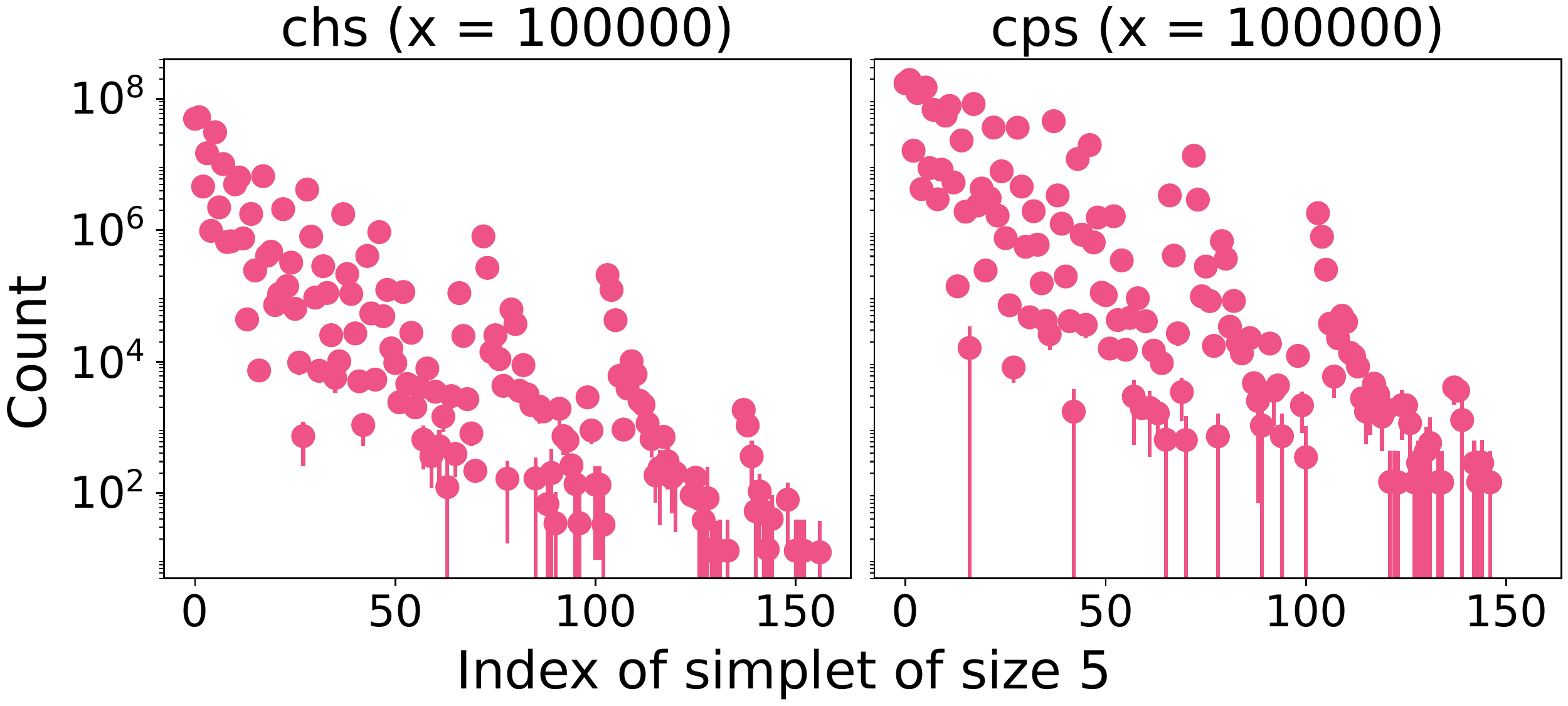} \\
    \caption{\underline{\smash{The estimated counts converge as the number of}} \underline{\smash{samples increases.}} For $k = 5$, we computed the estimated counts on the \textt{cps} (left) and \textt{chs} (right) datasets when the number of samples $x=100,000$.}
    \label{fig:conv5}
\end{figure}

In addition, we visualized how the count of each simplet converges to the exact count as the number of samples increases on the \coauthMAGGeology dataset.
As shown in Figure~\ref{fig:conv4}, the count of each simplet successfully converges to the actual count, and the standard deviation of the estimation for each simplet also decreases, as the number of samples increases. We additionally computed the mean values and the standard deviations of the estimated counts
when $k = 5$ on the \contacthighschool and \contactprimaryschool datasets.
As seen in Figure~\ref{fig:conv5}, the count of each simplet converges when the actual count is large enough, e.g., when we set the number of samples to $100,000$.

\begin{figure}[t]
    \vspace{-2mm}
    \centering
    \includegraphics[width=\linewidth]{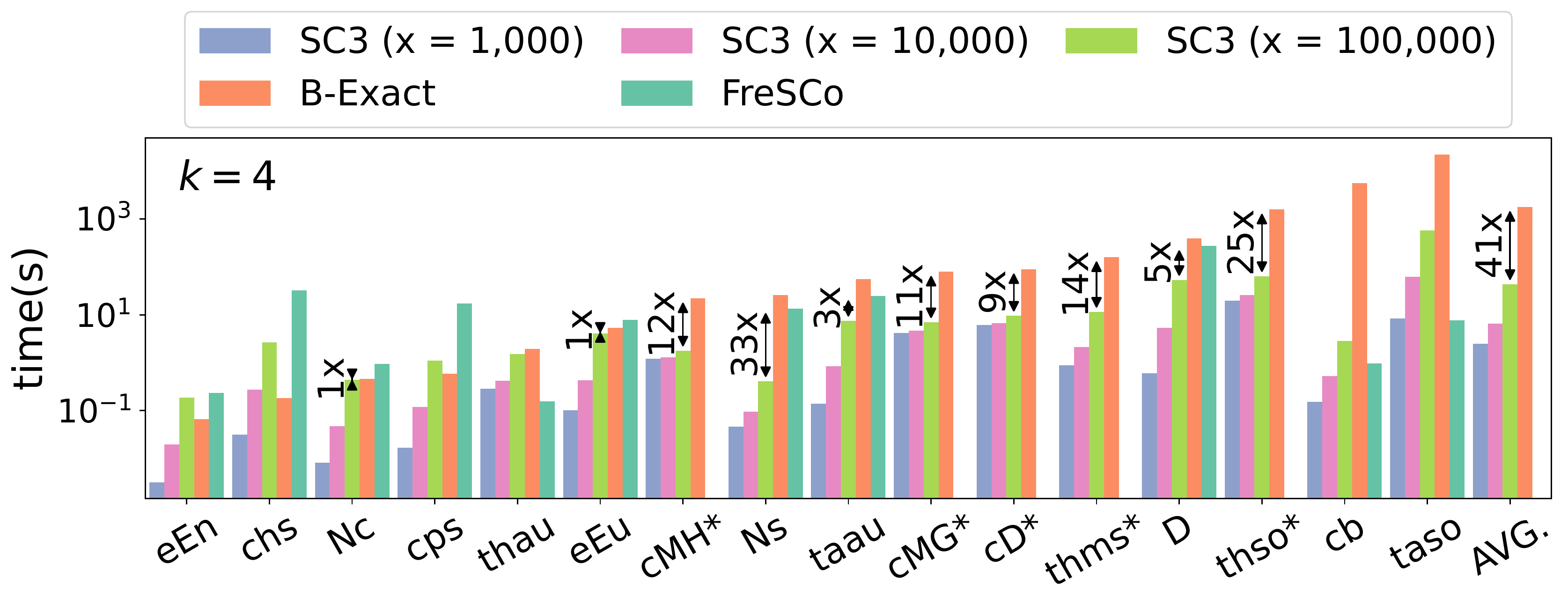}
    \includegraphics[width=\linewidth]{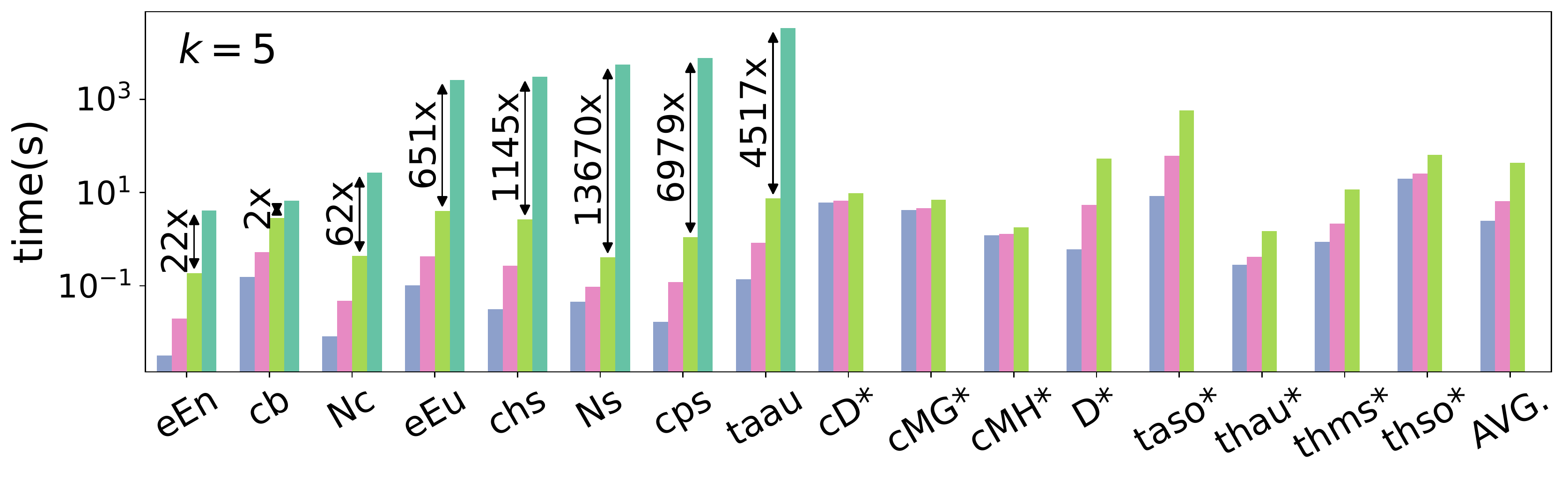} \\
    \vspace{-1mm}
    \caption{\underline{\smash{\ours is significantly faster than the baseline algo-}} \underline{\smash{rithms.}} \FreSCo ran out of time (> 10 hours) for some datasets, and \Benson can be used only for $k \leq 4$. %
    }
    \label{fig:time_compare}
\end{figure}

\subsection{\textbf{Q2. Scalability and Speed}}

We compared the running times of the considered algorithms.
Specifically, we measured the running time of \ours with the number of samples $x \in \{10^3, 10^4, 10^5\}$. For the baseline algorithms, we used \Benson for $k = 4$ and \FreSCo for $k = 4$ and $k = 5$.
As seen in Figure \ref{fig:time_compare}, \ours is fastest on $11$ (out of $16$ datasets) and is second fastest on the others. On average, \ours is $41\times$ faster than \Benson.

\begin{figure}[t]
    \vspace{-2.5mm}
    \centering
    \includegraphics[width=0.968\linewidth]{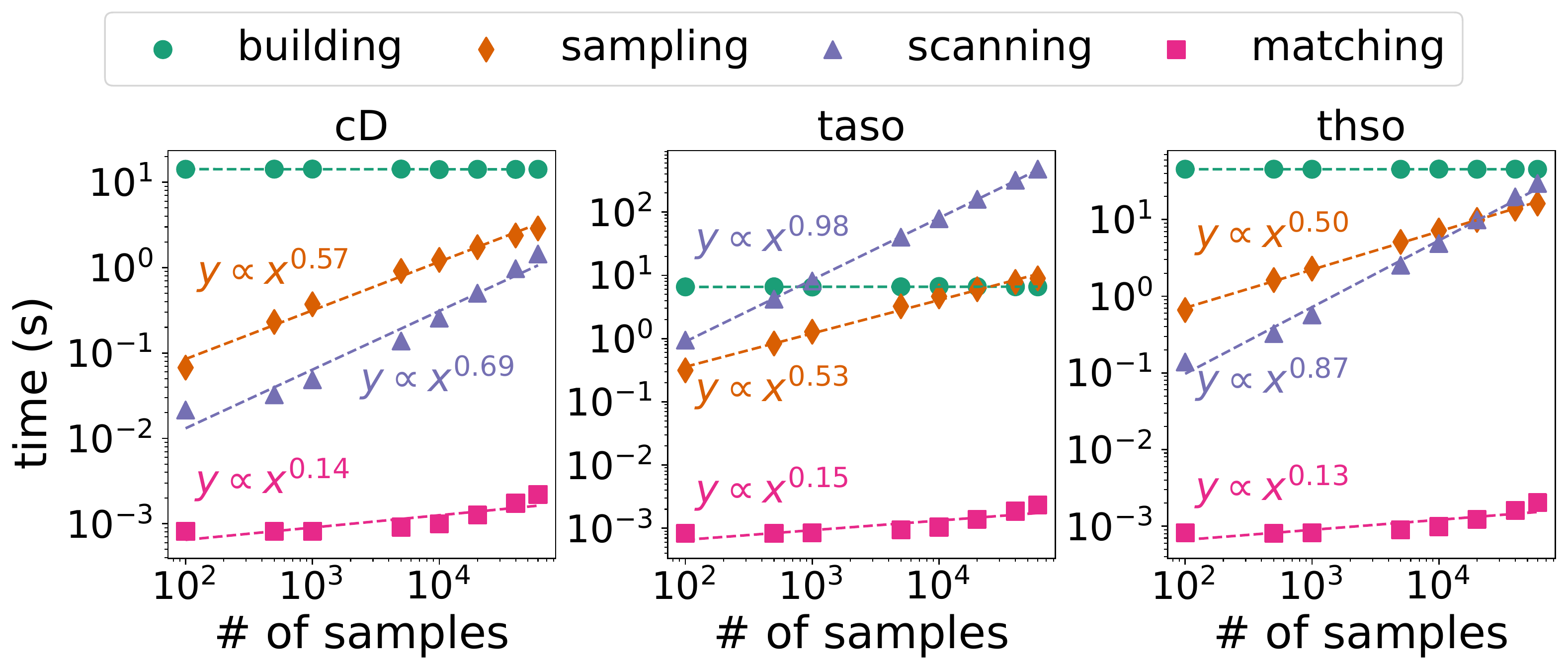} \\
    \vspace{-1mm}
     \caption{\underline{\smash{Each step of \ours is scalable.}} The running times of sampling, scanning, and matching steps increase sub-linearly as the number of samples increases.}
     \label{fig:tstep}
\end{figure}

To evaluate the scalability, we additionally measured the running time of each step of \ours on the \coauthDBLP, \tagsstackoverflow, and \threadsstackoverflow datasets.
We measured the running times with different numbers of samples.
As shown in Figure~\ref{fig:tstep}, the running times of sampling, scanning, and matching steps increase sub-linearly as the number of samples increases.
Note that the running time of the building step is constant that does not depend on the number of samples.

 \begin{figure*}[t]
    \vspace{-2mm}
    \begin{subfigure}[b]{\textwidth}
    \centering
        \includegraphics[width=0.944\linewidth]{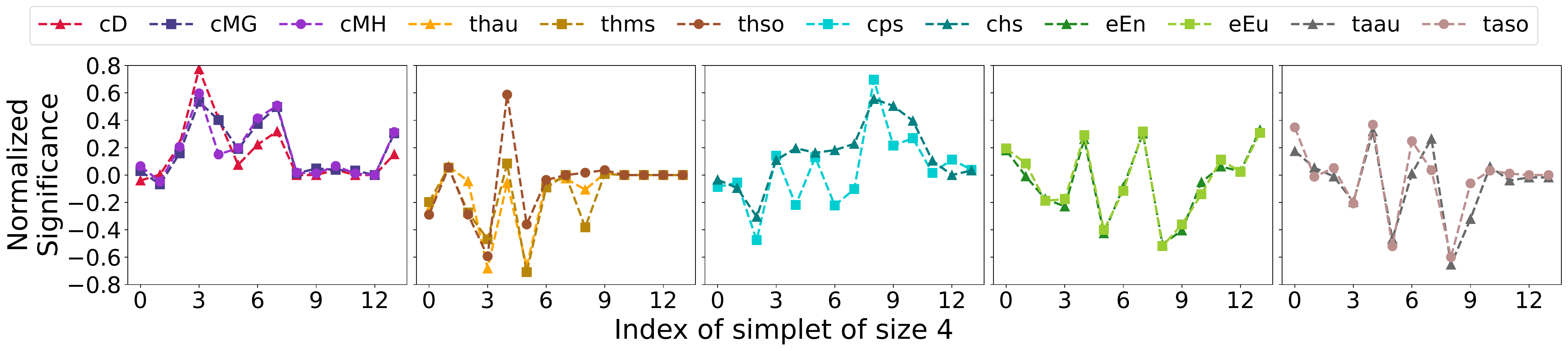}
    \end{subfigure}
    \begin{subfigure}[b]{\textwidth}
    \centering
        \includegraphics[width=0.944\linewidth]{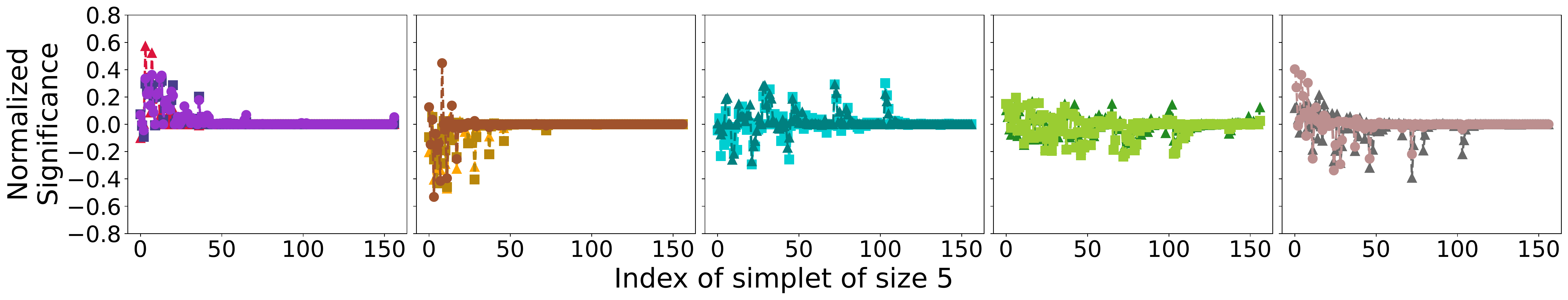}
    \end{subfigure} \\
     \caption{\underline{\smash{The SC datasets from the same domain show similar CPs.}}
     We report the characteristic profile (CP) of each dataset based on the counts of simplets of sizes $4$ (top) and $5$ (bottom) estimated by \ours.    %
     }
     \label{fig:sig}
\end{figure*}
 
 \begin{figure*}[t]
     \vspace{-2mm}
    \begin{subfigure}[b]{0.245\linewidth}
    \centering
        \includegraphics[width=\linewidth]{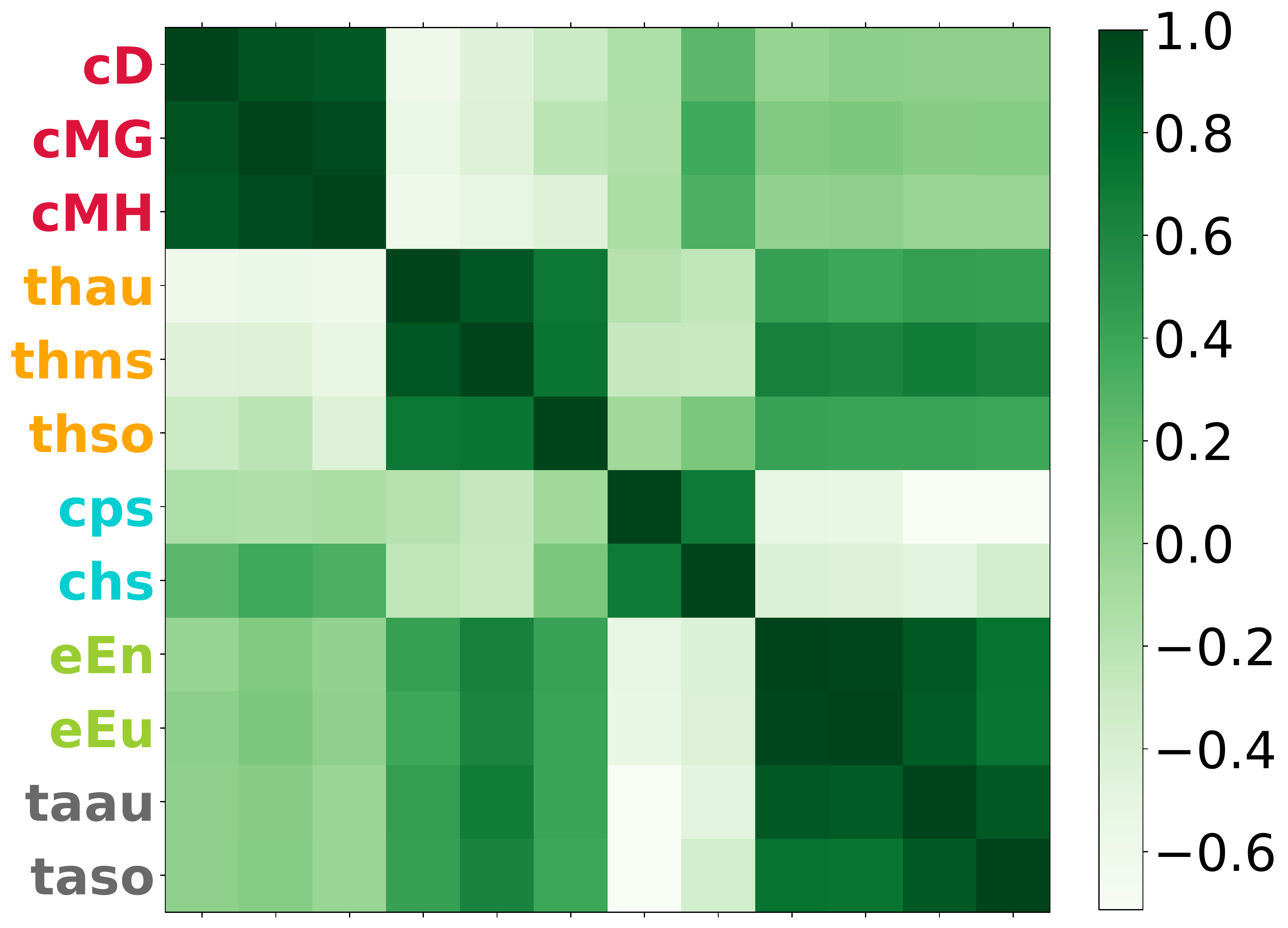}
        \caption{\ours with $k=4$}
        \label{fig:cp4}
    \end{subfigure}
    \begin{subfigure}[b]{0.245\linewidth}
    \centering
        \includegraphics[width=\linewidth]{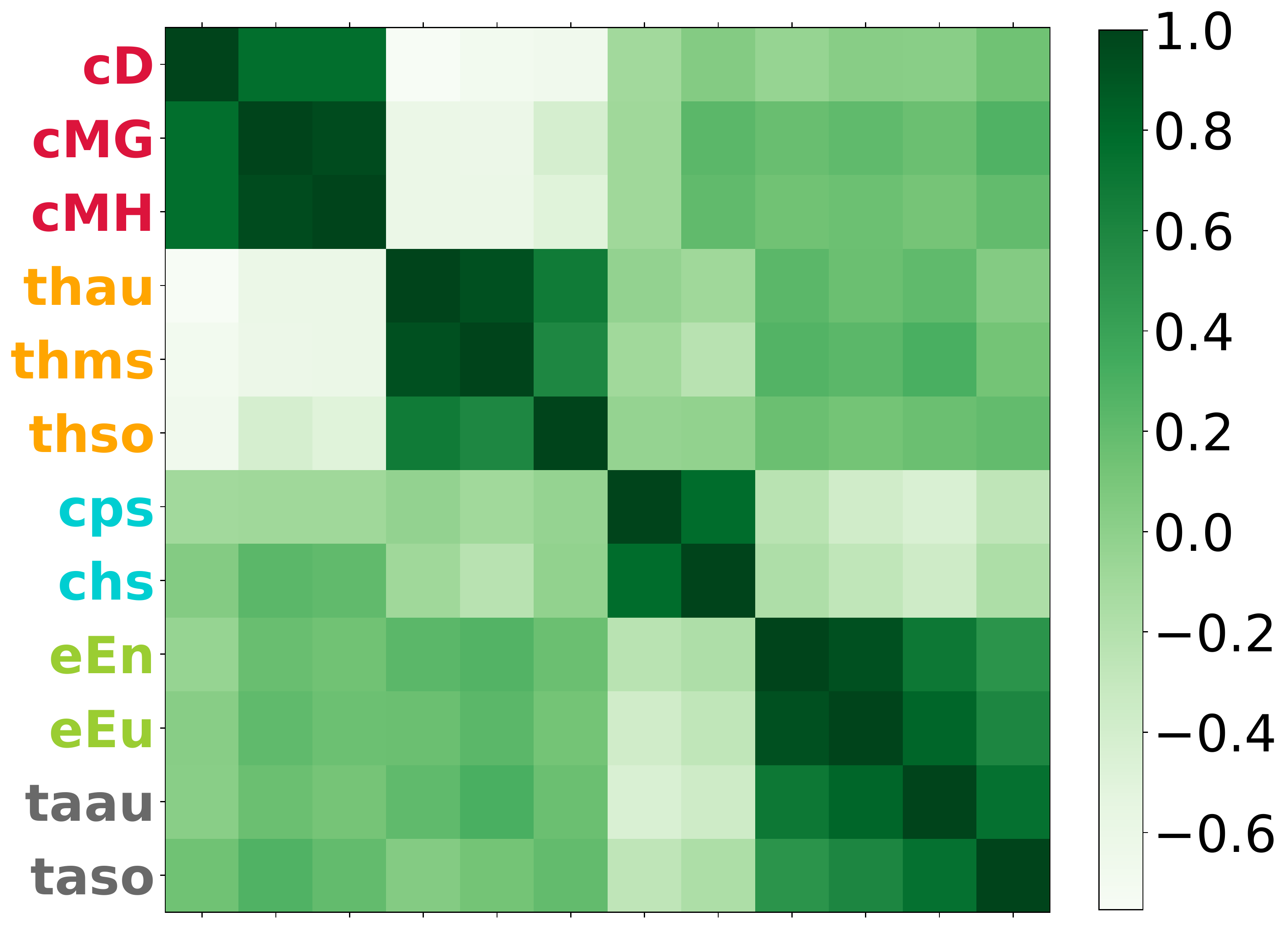}
        \caption{\ours with $k=5$}
        \label{fig:cp5}
    \end{subfigure}
    \begin{subfigure}[b]{0.245\linewidth}
    \centering
        \includegraphics[width=\linewidth]{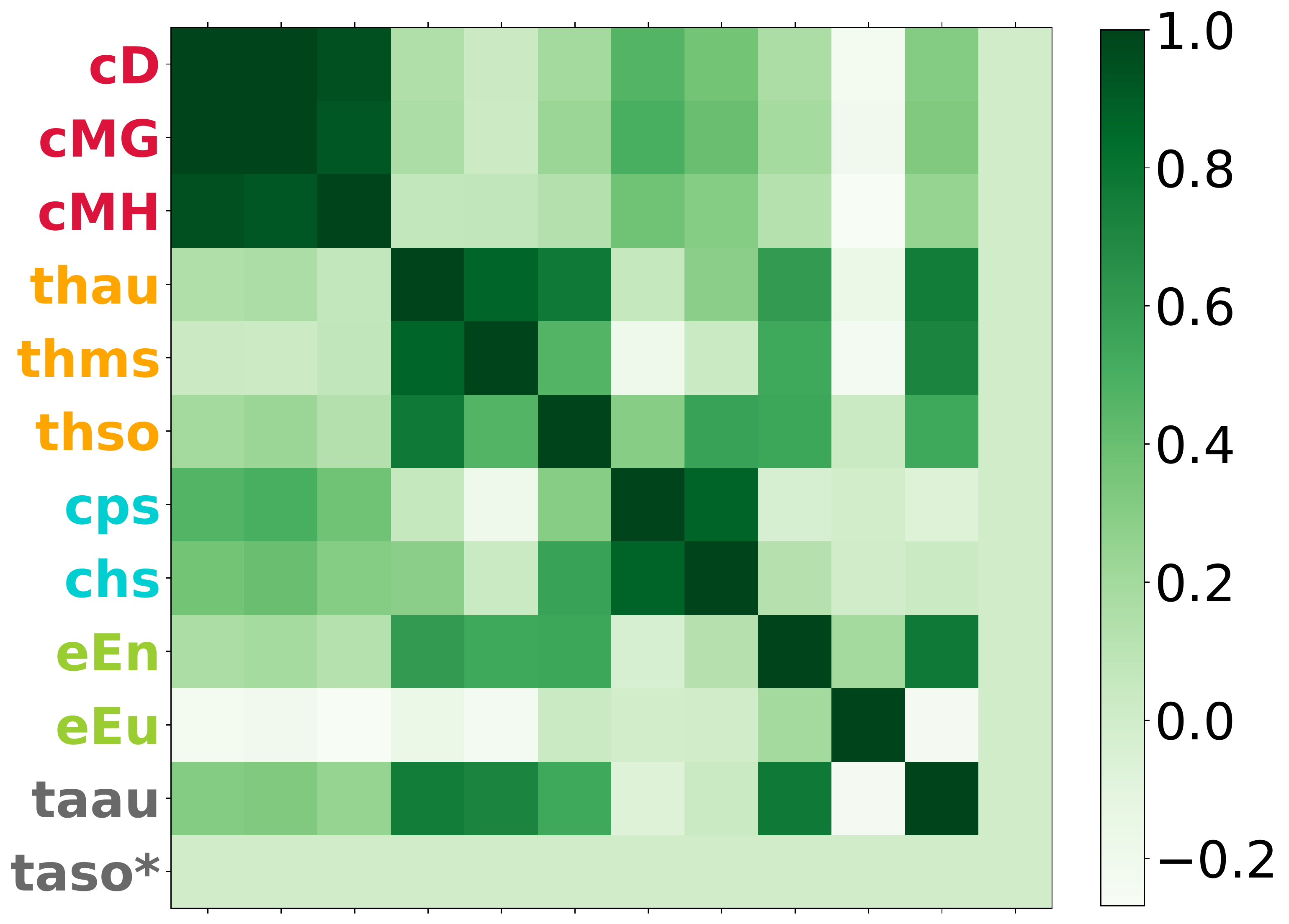}
        \caption{\Benson with $k=4$}
        \label{fig:cpb}
    \end{subfigure}
    \begin{subfigure}[b]{0.24\linewidth}
    \centering
        \includegraphics[width=\linewidth]{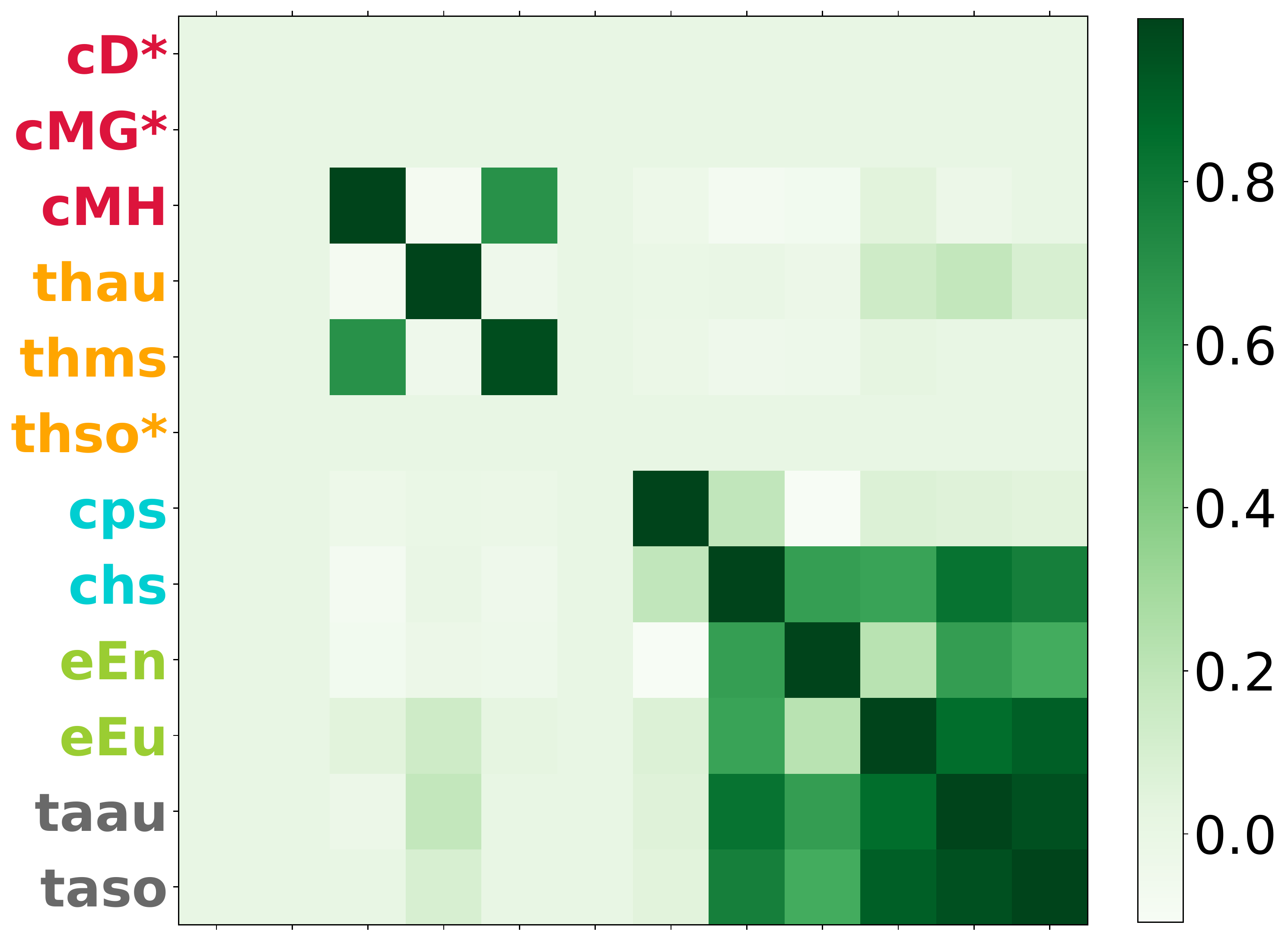}
        \caption{\FreSCo with $k=4$}
        \label{fig:cpf}
    \end{subfigure} \\
     \caption{
     \underline{\smash{The domains of the datasets are better distinguished when the CPs are based on the outputs of \ours.}}
     For each considered algorithm, we report the similarity between the CPs of each pair of datasets.
     The detailed numerical results are in Table~\ref{tab:km}.
     An asterisk (*) indicates that \Benson and \FreSCo ran out of time ($>$ 10 hours) on the corresponding dataset.}
\end{figure*}

\subsection{Q3. Characterization Power across Domains}\label{sec:exp_net}
To analyze the characterization power of the counts of simplets obtained by \ours, 
we computed the characteristic profile for each dataset.
We first analyzed SCs within the same domain using CPs and demonstrated them in Figure~\ref{fig:sig}.
Compared to the random SCs generated by the null model, the original SCs in the \textt{coauth-*} domain (\coauthDBLP, \coauthMAGGeology, \coauthMAGGeology) commonly contain more $\mathcal{S}^4_3$ and $\mathcal{S}^4_7$ with noticeable differences.
In the \textt{threads-*} domain (\threadsaskubuntu, \threadsmathsx, \threadsstackoverflow), two simplets $\mathcal{S}^4_3$ and $\mathcal{S}^4_5$, which do not contain any size-$3$ simplex, are less in the original SCs, compared to the random SCs generated by the null model.
For the \textt{contact-*} domain (\contacthighschool, \contactprimaryschool), 
the numbers of $\mathcal{S}_8^4$ (4-clique) in the real-world SCs are much higher than those in the random SCs generated by the null model.
In the \textt{email-*} domain (\emailEu, \emailEnron), two  
simplets $\mathcal{S}_5^4$, $\mathcal{S}_8^4$, which do not contain any size-$3$ simplex, are relatively less, while the other two simplets $\mathcal{S}_4^4$ and $\mathcal{S}_7^4$ those containing one or more size-$3$ simplices are relatively more, compared to the random SCs generated by the null model.
The \textt{email} domain and the \textt{tags} domain (\tagsstackoverflow, \tagsaskubuntu) are similar to each other when $k = 4$, except that the normalized significance of $\mathcal{S}_{13}^4$ of is noticeably different in the two domains.

\begin{table}[t]
\vspace{-2mm}
\centering
\caption{The result of k-means++ (i.e., the assignment of each dataset in clusters C1 and C5)  for 10 trials. When $k=5$, it always succeeded in distinguishing domains, but when $k=4$ it succeeded only once in ten trials.}
\scalebox{0.84}{
\centering
\setlength{\tabcolsep}{2.4pt}
\hspace{-1.5mm}
\begin{tabular}{c|ccc|ccc|cc|cc|cc|c}
\toprule
\multirow{2}{*}{k} & \multicolumn{3}{c|}{\texttt{coauth-*}} & \multicolumn{3}{c|}{\texttt{threads-*}} & \multicolumn{2}{c|}{\texttt{contact-*}} & \multicolumn{2}{c|}{\texttt{email-*}} & \multicolumn{2}{c|}{\texttt{tags-*}} & \multirow{2}{*}{trials}\\
\cline{2-13}
   & \coauthDBLP & \coauthMAGGeology & \coauthMAGHistory & \threadsaskubuntu & \threadsmathsx & \threadsstackoverflow & \contactprimaryschool & \contacthighschool & \emailEnron & \emailEu & \tagsaskubuntu & \tagsstackoverflow & \\
\hline
  $5$ & C1 & C1 & C1 & C2 & C2 & C2 & C3 & C3 & C4 & C4 & C5 & C5 &  10/10 \\
  \midrule
  \multirow{2}{*}{$4$} & C1 & C1 & C1 & C2 & C2 & C3 & C4 & C4 & C5 & C5 & C5 & C5 & 9/10 \\
    & C1 & C1 & C1 & C2 & C2 & C2 & C3 & C3 & C4 & C4 & C5 & C5 & 1/10\\
\bottomrule %

\end{tabular}
}
\label{tab:km}
\end{table}

To further evaluate the characteristic power across domains, 
we obtained $CP$ vectors for the considered algorithms.
Let $CP^4$, $CP^5$, $CP^B$ and $CP^F$ denote the CPs obtained from \ours with $k=4$, \ours with $k=5$, \Benson with $k=4$, and \FreSCo with $k = 4$, respectively.
For each type of CP, we computed the cosine similarity between each pair of the datasets.
Specifically, for $CP_B$, we computed CP using 4-node configurations described in the original paper instead of simplets;
for $CP_F$, 
we use the lowest $\tau \in \{10^5, 10^4, 10^3, 10^2, 10, 1\}$ such that \FreSCo terminates in $10$ hours.
Since \FreSCo computes the supports instead of the counts of simplets, 
we computed $CP^F$ using the ratio of the supports (instead of the counts used for \ours and \Benson).

As seen in Figures \ref{fig:cp4} and \ref{fig:cp5}, both $CP^4$ and $CP^5$ show strong characterization power, and the superiority of \ours w.r.t the characterization power becomes clearer when $k=5$.
Both competitors failed to clearly distinguish the SCs in different domains, which indicates their poor characterization power, as shown in Figures \ref{fig:cpb} and \ref{fig:cpf}.
We further embedded the CP of each SC in the Euclidean space and performed k-means++ clustering~\cite{arthur2006k} for 10 independent trials.
In Table~\ref{tab:km}, we report the clustering results using the CP computed from the counts obtained by \ours,
where the clustering results were perfect when $k = 5$,
while for $k = 4$, it often failed to cluster the SCs properly. 
The clustering results suggest that the $CP$ vectors w.r.t the counts of simplets can be applied to further downstream tasks, such as SC clustering or classification.

\smallsection{Extra results}
Simplets and SC3 can also be used to characterize nodes, and the results can be used as input for node-level tasks.
We present an experiment on node classification in \cite{appendix}.

\section{Conclusions}
\label{sec:conclusion}
In this work, we study the problem of counting simplets and develop \ours, a color-coding-based algorithm for it. We also show that the counts of simplets are effective in the characterization of real-world simplicial complexes (SCs).
Our contributions are three-fold:
\begin{itemize}[leftmargin=*]
    \item \textbf{New Problem.} To the best of our knowledge, we are the first to formulate and study the problem of directly counting simplets in a given SC, especially for the simplets beyond four nodes.
    \item \textbf{Accurate and Fast Algorithm.} \ours is orders of magnitude faster than its competitors. %
    Empirically, its running time is sub-linear w.r.t the number of samples.
    As a result, \ours succeeds in estimating the count of every simplet of size $4$ and $5$ in large SCs, and the result is accurate with theoretical guarantees.
    \item \textbf{Characterization of Real-world SCs.} We demonstrated that the output of \ours can be used to characterize SCs. %
    Especially, the characteristic profiles (CPs) based on the count of simplets of size $5$ obtained by \ours better distinguish the 
    domains of real-world SCs than the CPs from its competitors.

\end{itemize}
For reproducibility, we make the code and datasets available at \cite{appendix}.

\vspace{1mm}
\smallsection{Acknowledgements}
This work was supported by Institute of Information \& Communications Technology Planning \& Evaluation (IITP) grant funded by the Korea government (MSIT) (No. 2022-0-00871, Development of AI Autonomy and Knowledge Enhancement for AI Agent Collaboration) (No. 2019-0-00075, Artificial Intelligence Graduate School Program (KAIST)).

\bibliographystyle{ACM-Reference-Format}
\bibliography{bib}

\appendix

\section{Proofs}\label{sec:pfs_A}

\begin{proof}[Proof of Lemma~\ref{lem:build_sample}] \label{prof:build_sample}
    See~\cite{bressan2018motif} (esp. Theorem 5.1), 
    combined with the fact that a subcomplex of an SC $\mathcal{K}$ induced on a node set $V'$ is connected if and only if 
    the induced subgraph of its primal graph on the same node set is connected.
\end{proof}

\begin{proof}[Proof of Lemma~\ref{lem:scan}] \label{prof:scan}
    For each $V_T \in T_{ct}$, the number of $\sigma$'s to check is $O(|M(\mathcal{K})|)$, and
    maintaining the set of all the maximal sets takes $O(|M(\mathcal{K})||M(\mathcal{K}[V_T])|)$ time~\cite{yellin1992algorithms}.
    The space complexity is straightforward,
    and the correctness immediately follows the definition of maximal simplices.
\end{proof}

\begin{proof}[Proof of Lemma~\ref{lem:match}] \label{prof:match}
    Building the permutation-invariant maps takes $O(|M(\mathcal{S})|)$ for each bijection $\phi$ and for each $\mathcal{S} \in \mathcal{S}^k$.
    Enumerating the whole $T_{ct}$ and accumulating the estimated counts takes $O(|T_{ct}|)$ times.
    The space complexity is straightforward.
\end{proof}

\begin{proof}[Proof of Theorem~\ref{thm:unbiasedness}]
    By Lemma~\ref{lem:build_sample}, for each $\mathcal{S} \in \mathcal{S}^k$,
    let $V_\mathcal{S} \subseteq \binom{V}{k}$ with $|V_\mathcal{S}| = N_{oc}(\mathcal{S})$ denote the set of occurrences of $\mathcal{S}$.
    Each connected $V_T \in V_\mathcal{S}$ is colorful with probability $k! / k^k$,
    and at each iteration of sampling, $V_T$ is sampled with probability $n_{st}(G_{\mathcal{K}}[V_T]) / N_{ct}$.
    By the correctness of all the steps (Lemmas~\ref{lem:build_sample}-\ref{lem:match}),
    the corresponding simplet of each sampled $V_T$ is correctly found.
    Therefore, $\mathbb{E}[\tilde{N}_{oc}(S)] = \sum_{V_T \in V_\mathcal{S}} |T_{ct}| \frac{k!}{k^k} \frac{n_{st}(G_{\mathcal{K}}[V_T])}{N_{ct}} \frac{1}{n_{st}(G_{\mathcal{K}}[V_T])} \frac{N_{ct}}{|T_{ct}|} \frac{k^k}{k!} = \sum_{V_T \in V_\mathcal{S}} 1 = N_{oc}(\mathcal{S})$.
\end{proof}

\begin{proof}[Proof of Theorem~\ref{thm:convergence}]
    It also relies on the correctness of all the steps (Lemmas~\ref{lem:build_sample}-\ref{lem:match}).
    The statement 
    for one trial is essentially follows Chebyshev's inequality where \[\sigma^2=Var[\tilde{N}_{oc}(\mathcal{S})]=\frac{N_{oc}(\mathcal{S})}{x} \left(\frac{N_{ct}\cdot k^k/k!}{n_{st}(\mathcal{S})} - 1\right).\] 
    In addition, we have a Chebyshev bound (notations borrowed from Corollary 5.5 in~\cite{bressan2018motif}) of the probability that a uniformly-randomly-drawn colorful simplet corresponds to $S$: the probability is in $\mu_S \pm 2\epsilon/(1+\epsilon)$ with probability $1-\Omega((x\epsilon)^{-2})$.
\end{proof}

\begin{proof}[Proof of Theorem~\ref{thm:complexities}]
    By Lemmas~\ref{lem:build_sample}-\ref{lem:match}, the total time complexity is
    $O(c^k |E_G| + xk + |M(\mathcal{K})| \hat{M}_{ct} + k!\hat{M}_k + x) = O(c^k |E_G| + x|M(\mathcal{K})|^2 + k! \hat{M}_k)$,
    where $E_G=E\cap \binom{V}{2}$, and we have used $\hat{M}_{ct} = O(x |M(\mathcal{K})|)$ and $k = O(M(\mathcal{K}))$.
    and the total space complexity is 
    $O(c^k|E_G| + \hat{M}_{ct} + \hat{M}_k + \hat{M}_{ct}) = O(c^k|E_G| + x |M(\mathcal{K})| + \hat{M}_k)$. \footnote{The terms can be simplified by using $\max(|E_G|,M(\mathcal{K}))\leq|E|$ and regarding $x$ as a constant, which gives $O(c^k|E| + |E|^2 + k!\hat{M}_k)$ and $O(c^k|E| + \hat{M}_k)$.}
\end{proof}

\begin{table}[t] 
    \centering
    \caption{The outputs of \Benson, \FreSCo, and \ours for $k=4$ on the \textt{Ns} dataset.
    We run \ours with $x = 100,000$ samples and report the means over $5$ independent trials.
    For \FreSCo, we set the threshold $\tau$ to $1$ to obtain the most accurate output, and set a time limit of $10$ hours.}

    \begin{adjustbox}{max width=1\linewidth}
    \begin{tabular}{c|r|r|r}
        \toprule
        \textbf{index of $\mathcal{S}^4$} &  \textbf{\Benson} & \textbf{\ours} &\textbf{\FreSCo} \\
        \midrule
        0 & - & 953999597.1 & 2973 \\
        1 & - & 1481594205.7 & 3096 \\ 
        2 & - &  19876052.5 &  3024 \\ 
        3 & 528575876  & 527726043.0 & 2964\\ 
        4 & 278014263  & 279769791.8 & 2955 \\
        5 & 68051012  & 68128876.7 & 2940 \\
        6 & 61794552  & 61601934.2 & 2935 \\ 
        7 & 17027710 & 17451818.3 & 2917 \\ 
        8 & 7538854  & 7648915.7 & 2908 \\ 
        9 & 10806231  & 10907215.3 & 2903 \\ 
        10 & 6040967  & 6102560.0 & 2896 \\ 
        11 & 1577597  & 167242.9 & 2885 \\
        12 & 168752  & 139459.2 & 2885 \\
        13 & - & 3290324.8 & 2885 \\
        \bottomrule
    \end{tabular}
        \label{tab:fresco}
    \end{adjustbox}
\end{table}

\begin{algorithm}[t]
    \caption{Simplet expansion from graphlets}
    \label{algo:simplet}
    \SetKwInput{KwInput}{Input}
    \SetKwInput{KwOutput}{Output}
    
    \KwInput{(1) $k$: the considered size of graphlets and simplets\\
    \quad\quad\quad (2) $\mathcal{G}^k$: the set of all the graphlets of size $k$}
    \KwOutput{$\mathcal{S}_k$: the set of all the simplets of size $k$}
    
    \SetKwProg{Fn}{Function}{:}{\KwRet}
    $\mathcal{S}_k$ $\leftarrow$ $\emptyset$ \Comment*[f]{Initialization} \\
    \Fn{\textnormal{Expand($i, C, S$)}  }{
        \If{$i \leq 2$ \label{algo:simplet:term} }{
            $\mathcal{S}_k$ $\leftarrow$ $\mathcal{S}_k \cup \{S\}$ \label{algo:simplet:simplet} \\
            \Return \label{algo:simplet:line2}\\
        }
        $T$ $\leftarrow$ $\binom{C}{i} \setminus S$;
        $Q$ $\leftarrow$ $\emptyset$
        \label{algo:simplet:T} \\
        \ForEach{$\Sigma \subseteq T$}{\label{algo:simplet:cases}
            $S'$ $\leftarrow$ $S \cup \left(\bigcup_{I\in \Sigma} 2^I\right)$\\
            \If{$\nexists q \in Q$ s.t. $S' \simeq q$ \label{algo:simplet:Q}}{
                $Q$ $\leftarrow$ $Q \cup \{S'\}$ \label{algo:simplet:Q2}\\
                $\textnormal{Expand}(i-1, C, S')$\\
            }
        }
    }
    \For{$\mathcal{G} = (V_\mathcal{G} = [k], E_\mathcal{G}) \in \mathcal{G}^k$}{
        $C_\mathcal{G} \leftarrow \{C \subseteq [k] : |C| > 2 \wedge \binom{C}{2} \subseteq E_\mathcal{G} \}$ \label{algo:simplet:init} \Comment*[f]{Cliques} \\
        $\textnormal{Expand}(k, C_\mathcal{G}, \{\emptyset\} \cup \binom{[k]}{1} \cup E_\mathcal{G})$\label{algo:simplet:init2} \\
    }
 \Return $S_k$ 
\end{algorithm}

\begin{algorithm}[t]
    \small
    \caption{\ours-build}\label{algo:build}
    \SetKwInput{KwInput}{Input}
    \SetKwInput{KwOutput}{Output}
    \SetKw{Continue}{continue}
    \KwInput{(1) $k$: the considered size of simplets \\
    \quad\quad\quad (2) $G_\mathcal{K}=(V_G, E_G)$: the primal graph of the input SC $\mathcal{K}$}
    \KwOutput{(1) $C(v, \mathcal{T}, S), \forall v \in V, \mathcal{T}, S$: the number of occurrences of \\
    \quad\quad\quad\quad\ \ $\mathcal{T}$ colored by $S$ rooted at $v$ \\
    \quad\quad\quad\ \ (2) $N_{ct}$: the total count of occurrences of size-$k$ colorful\\
    \quad\quad\quad\quad\ \  treelets} 
    \ForEach{$v \in V$}{
        \ForEach{$\mathcal{T} \in \bigcup_{i=1}^{k} \mathcal{T}^i$}{
            \ForEach{$S \in \binom{[k]}{|\mathcal{T}|}$}{
                $C(v, \mathcal{T}, S) \leftarrow 0$ \\
            }
        }
    }
    \ForEach{$v \in V$}{
        $c(v) \leftarrow$ uniformly at random sampled in $[k]$ \label{algo:build:color}\\
        $C(v, (\{0\}, \emptyset), \{c(v)\}) \leftarrow 1$ \label{algo:build:init} \\
    }
    \For{$i = 2, \dots, k$ \label{algo:build:treesize}}{
        \ForEach{$v \in V$}{
            \If{$i = k$ and $c(v) \neq 0$}{ \label{algo:build:cond}
                \Continue
            }
            \ForEach{$\mathcal{T} \in \mathcal{T}^i$}{
                \ForEach{$S \in \binom{[k]}{i}$}{
                    $C(v, \mathcal{T}, S) \leftarrow \mathlarger{\frac{1}{d}}\mathlarger{\sum\limits}_{(u,v)\in E_{\mathcal{K}}} {\mathlarger{\sum\limits}_{S_1 \sqcup S_2 = S}} C(v, \mathcal{T}_1, S_1) \cdot C(u, \mathcal{T}_2, S_2)$\label{algo:build:equat}\\
                }
            }
        }   
    }
    $N_{ct} \leftarrow 0$ \\
    \For{$v \in V$}{
         \ForEach{$\mathcal{T} \in \mathcal{T}^k$}{
            $N_{ct} \leftarrow N_{ct} + C(v, \mathcal{T}, [k])$ \\
         }
    }
    \Return $C$, $N_{ct}$ \label{algo:build:end}
\end{algorithm}

\begin{algorithm}[t]
    \small
    \caption{\ours-sample \label{algo:sample}}
    \SetKwInput{KwInput}{Input}
    \SetKwInput{KwOutput}{Output}
    \KwInput{(1) $k$: the considered size of simplets \\
    \quad\quad\quad (2) $G_\mathcal{K}=(V_G, E_G)$: the primal graph of the input SC $\mathcal{K}$ \\
    \quad\quad\quad (3) $C(v, \mathcal{T}, S), \forall v \in V, \mathcal{T}, S$: the number of occurrences of \\
    \quad\quad\quad\quad\ \  $\mathcal{T}$ colored by $S$ rooted at $v$ \\
    \quad\quad\quad (4) $N_{ct}$: the total count of occurrences of size-$k$ \\
    \quad\quad\quad\quad\ \ colorful treelets \\
    \quad\quad\quad (5) $x$: the number of samples
    }
    \KwOutput{$T_{ct}$: the sampled occurrences of colorful treelets}
    
  \SetKwProg{Fn}{Function}{:}{\KwRet}
  \Fn{$\textnormal{Sample($v, \mathcal{T}, S$)}$}{
        \If{$|\mathcal{T}|=1$}{
        \Return{$\{v\}$}
    }
        decompose $\mathcal{T}$ with $\mathcal{T}_1$ and $\mathcal{T}_2$\\
        choose $u \in N(v), S_1 \in \binom{[k]}{|\mathcal{T}_1|}, S_2 \in \binom{[k]}{|\mathcal{T}_2|}$ from distribution $\Pr(u, S_1, S_2)\propto C(v, \mathcal{T}_1, S_1) \cdot C(u, \mathcal{T}_2, S_2)$\label{algo:sample:line3}\\
        \Return $\textnormal{Sample}(v, \mathcal{T}_1, S_1)\cup \textnormal{Sample}(u, \mathcal{T}_2, S_2)$
      }
    $T_{ct}=\emptyset$ \\
    \ForEach{$i \in [x]$}{
        Choose $v, \mathcal{T}$ from distribution $\Pr(v,\mathcal{T}) = C(v, \mathcal{T}, [k]) / N_{ct}$\label{algo:sample:line1}\\
        $V_T = \textnormal{Sample$(v, \mathcal{T}, [k])$}$\\
        $T_{ct} \leftarrow T_{ct} \cup \{V_T\}$
    }
     \Return $T_{ct}$ \label{algo:sample:end}
\end{algorithm}

\section{Baseline Algorithms}\label{app:base}
\smallsection{\Benson}
For each size-$4$ simplet, we show which 4-node configurations (see Table~7 in the appendix of~\cite{benson2018simplicial} for the details), if any, are corresponding to it:
    (1) $\mathcal{S}_4^{0}: \emptyset$;
    (2) $\mathcal{S}_4^{1}: \emptyset$;
    (3) $\mathcal{S}_4^{2}: \emptyset$;
    (4) $\mathcal{S}_4^{3}: \psi_0$;
    (5) $\mathcal{S}_4^{4}: \psi_1$, $\psi_2$;
    (6) $\mathcal{S}_4^{5}: \theta_{0,0}$;
    (7) $\mathcal{S}_4^{6}: \theta_{0,1}$, $\theta_{0,2}$;
    (8) $\mathcal{S}_4^{7}: \theta_{1,1}$, $\theta_{1,2}$, $\theta_{2,2}$;
    (9) $\mathcal{S}_4^{8}: \pi_{0,0,0,0}$;
    (10) $\mathcal{S}_4^{9}: \pi_{0,0,0,1}$, $\pi_{0,0,0,2}$;
    (11) $\mathcal{S}_4^{10}: \pi_{0,0,1,1}$, $\pi_{0,0,1,2}$, $\pi_{0,0,2,2}$;
    (12) $\mathcal{S}_4^{11}: \pi_{0,1,1,1}$, $\pi_{0,1,1,2}$, $\pi_{0,1,2,2}$, $\pi_{0,2,2,2}$;
    (13) $\mathcal{S}_4^{12}: \pi_{1,1,1,1}$, $\pi_{1,1,1,2}$, $\pi_{1,1,2,2}$, $\pi_{1,2,2,2}$, $\pi_{2,2,2,2}$;
    (14) $\mathcal{S}_4^{13}: \emptyset$.
We can see that the configurations in~\cite{benson2018simplicial} and the simplets are not one-to-one corresponded.

\smallsection{\FreSCo}
We provide the definition of \textit{support}, the surrogate measure used in~\cite{preti2022fresco}, which is essentially different from the count of simplets \textit{per se}, both theoretically and empirically.
\begin{definition}[support~\cite{preti2022fresco}]
Let the image set $I_P(v)$ of $v\in V_P$ of a simplet $P$ be the set of vertices in $\mathcal{K}$ that are mapped to $v$ by some isomorphism $\phi$, i.e., $I_P(v)=\{u\in \mathcal{K}\mid \exists \phi \textnormal{ s.t. } \phi(u)=v\}$. Then the support of $P$ in $\mathcal{K}$ is $SUP(P, \mathcal{K})=\min_{v\in V_P} |I_P(v)|$.
\end{definition}

In Table~\ref{tab:fresco}, for $k = 4$, we compare the outputs of \Benson,\footnote{For \Benson, for each simplet, we sum up the counts of the node configurations corresponding to the simplet.} \FreSCo, and \ours,
where we can see that the output of \ours is an accurate estimator of the exact count of simplets,
while the output of \FreSCo is hardly meaningful.

\section{Simplet expansion from graphlets}\label{sec:simplet_expansion}
In Algorithm~\ref{algo:simplet}, we provide a practical way to generate the set of all possible simplets of size $k$.
Essentially, we conduct an expansion from all the graphlets of size $k$ to all the simplets of size $k$.
We assume that all the graphlets of size $k$ are known beforehand,\footnote{Technically, they can be generated via isomorph-free exhaustive generation~\cite{mckay1998isomorph}.} and use them as the input of \ours.
Specifically, we exploit the fact that, for each $k$, there is a surjection from all the simplets of size $k$ to all the graphlets of size $k$, where each simplet is mapped to its primal graph. 

We first save all the cliques of size more than 2 for each graphlet $\mathcal{G} \in \mathcal{G}^k$ (line~\ref{algo:simplet:init}), the candidates to be open or closed (i.e., covered with a simplex of the same size). Then, starting from cliques of possible maximum size (line~\ref{algo:simplet:init2}) to edges  (line~\ref{algo:simplet:term}), we consider the  number of all cases ($\Sigma\in 2^T$ on line~\ref{algo:simplet:cases}) that a clique of the current size $i$ any of whose superset is not closed ($T$ on line~\ref{algo:simplet:T}) is either open or closed recursively. If $S'$ made by closing every $\Sigma\in 2^T$ is not isomorphic $\forall q\in Q$ (line~\ref{algo:simplet:Q}), then we put $S'$ into a simplet set $Q$ (line~\ref{algo:simplet:Q2}) and keep expansion until there is no clique to be closed. Finally, when reaching the terminate condition, we put the simplet $S$ to the set of simplets $\mathcal{S}_k$ (line~\ref{algo:simplet:simplet}).

\section{A toy example for \ours}\label{app:example}
\begin{figure}[t]
    \centering
    \includegraphics[width=0.88\linewidth]{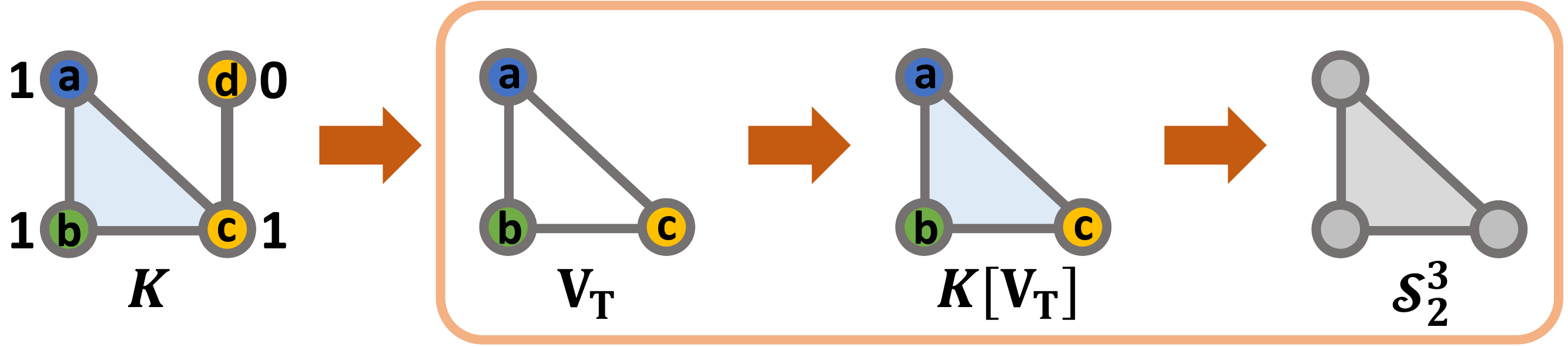} \\
    \caption{The subfigures correspond to the building, sampling, scanning, and matching steps in order. }
    \label{fig:toy_ex}
\end{figure}

See Figure~\ref{fig:toy_ex} for the entire procedure.
For a given SC of order 4 and $k=3$, we aim to count $N_{oc}(\mathcal{S}_i^3; \mathcal{K})$ for each $i\in [s_3]$. In the building step, we count the number of colorful treelets rooted on each node (in \textbf{bold} next to each node) after coloring every node with three colors uniformly at random. In the sampling step, $V_T=\{a, b, c\}$ is sampled with probability 1, the only case of a size-3 colorful tree. In the scanning step, an induced subcomplex on $V_T$ is found and matched to an isomorphic simplet ($\mathcal{S}^3_2$) in the final step. The three processes except for the building step are repeated. 

\begin{table}[t]
    \vspace{-2mm}
    \centering
    \caption{The Basic statistics of the real-world datasets.} %
    \label{tab:datasets}
    \scalebox{0.85}{
    \begin{tabular}{l|l|r|r}
        \toprule
        \textbf{Dataset $\mathcal{K}=(V, E)$} &\textbf{Abbreviation} & \textbf{$|V|$} & \textbf{$|M(\mathcal{K})|$} \\
        \midrule
        coauth-DBLP~\cite{benson2018simplicial} & \textt{cD} & $1,924,991$ & $1,730,664$ \\ %
        coauth-MAG-Geology~\cite{sinha2015overview, benson2018simplicial} & \textt{cMG} & $1,256,385$ & $925,027$ \\ %
        coauth-MAG-History~\cite{sinha2015overview, benson2018simplicial} & \textt{cMH} & $1,014,734$ & $774,495$ \\ %
        congress-bills~\cite{benson2018simplicial, fowler2006connecting, fowler2006legislative} & \textt{cb} & $1,718$ & $48,898$ \\ %
        contact-high-school~\cite{benson2018simplicial, mastrandrea2015contact} & \textt{chs} & $327$ & $4,862$ \\ %
        contact-primary-school~\cite{benson2018simplicial, stehle2011high} & \textt{cps} & $242$ & $8,010$ \\ %
        DAWN~\cite{benson2018simplicial} & \textt{D} & $2,558$ & $72,421$ \\ %
        email-Eu~\cite{benson2018simplicial, yin2017local, leskovec2007graph} & \textt{eEu} & $998$ & $8,102$ \\ %
        email-Enron~\cite{benson2018simplicial} & \textt{eEn} & $143$ & $433$ \\  %
        NDC-classes~\cite{benson2018simplicial} & \textt{Nc} & $1,161$ & $563$ \\ %
        NDC-substances~\cite{benson2018simplicial} & \textt{Ns} & $5,311$ & $6,555$ \\ %
        tags-ask-ubuntu~\cite{benson2018simplicial} & \textt{taau} & $3,029$ & $95,639$ \\ %
        tags-stack-overflow~\cite{benson2018simplicial} & \textt{taso} & $49,998$ & $3,781,574$ \\ %
        threads-ask-ubuntu~\cite{benson2018simplicial} & \textt{thau} & $125,602$ & $149,025$ \\ %
        threads-math-sx~\cite{benson2018simplicial} & \textt{thms} & $176,445$ & $519,573$ \\ %
        threads-stack-overflow~\cite{benson2018simplicial} & \textt{thso} & $2,675,955$ & $8,694,667$ \\ %
        \bottomrule
    \end{tabular}
    }
\end{table}

\section{Algorithmic Details of \ours}\label{sec:alg_A}
In Algorithms~\ref{algo:build} and \ref{algo:sample}, we provide the detailed processes of the building and sampling steps.

\section{Datasets}\label{sec:data}
We provide the basic statistics of the datasets in Table \ref{tab:datasets}.

\end{document}


\begin{table*}[t!]
\centering
\LARGE{Characterization of Simplicial Complexes by Counting Simplets Beyond Four Nodes \textcolor{red}{(Supplementary Document)}} \\
\textcolor{red}{{\fontsize{11}{100} \selectfont NOTE: If a preview does not appear properly, please download this file.}}
\end{table*}

\title{\vspace{-25mm}}
\maketitle

\section{Application~\rome{2}--- node classification (Related to Section 6.3)}

\begin{table}[t]
    \centering
    \caption{The training time per epoch on all datasets. We report the means and standard errors of training time on each dataset.}
    \begin{tabular}{c|c}
    \toprule
        Dataset & \multirow{2}{*}{Training time}\\
        (Hidden Dimension) & \\
    \midrule
        \texttt{email} (30) & 0.19 $\pm$ 0.010 \\
        \texttt{nyc} (30) & 0.21 $\pm$ 0.004 \\
        \texttt{tky} (30) & 0.32 $\pm$ 0.005 \\
        \texttt{kasandr} (60) & 1.93 $\pm$ 0.005 \\
        \texttt{threads} (60) & 5.49 $\pm$ 0.012 \\
        \texttt{twitch} (90) & 566.82 $\pm$ 3.308 \\
        \texttt{nips} (50) & 6.31 $\pm$ 0.081 \\
        \texttt{enron} (90) & 80.69 $\pm$ 0.266 \\
        \texttt{3-gram} (90) & 27.19 $\pm$ 0.089 \\
        \texttt{4-gram} (90) & 41.09 $\pm$ 0.785 \\
    \bottomrule
    \end{tabular}
    \label{tab:time_per_epoch}
\end{table}

\bibliographystyle{IEEEtran}